\documentclass[notitlepage,preprintnumbers,prd,longbibliography,showpacs,nofootinbib]{revtex4-1}
\usepackage{amsmath,amssymb}
\usepackage{graphics,color}
\usepackage{caption,subcaption}
\usepackage{epsfig}
\usepackage{dcolumn}
\usepackage{multirow}
\usepackage{bm}
\usepackage{verbatim}

\usepackage[colorlinks=true,
            urlcolor=blue,
            linkcolor=blue,
            citecolor=blue,
            bookmarks=true,
            bookmarksopen=true,
            bookmarksnumbered=true]{hyperref}

\begin{document}
\title{Doubly heavy tetraquark states in a mass splitting model}
\author{Shi-Yuan Li$^1$}
\author{Yan-Rui Liu$^1$}\email{yrliu@sdu.edu.cn}
\author{Zi-Long Man$^1$}\email{manzilong@mail.sdu.edu.cn}
\author{Zong-Guo Si$^1$}
\author{Jing Wu$^2$}\email{wujing18@sdjzu.edu.cn}

\affiliation{$^1$School of Physics, Shandong University, Jinan, Shandong 250100, China\\
$^2$School of Science, Shandong Jianzhu University, Jinan 250101, China
}

\date{\today}

\begin{abstract}
Treating the $X(4140)$ as a compact $J^{PC}=1^{++}$ $cs\bar{c}\bar{s}$ state and using its mass as a reference scale, we systematically estimate the masses of doubly heavy tetraquark states $QQ\bar{q}\bar{q}$ where $Q=c,b$ and $q=u,d,s$. Their decay properties are studied with a simple rearrangement scheme. Based on our results, the lowest $I(J^P)=0(1^+)$ $bb\bar{n}\bar{n}$ state is a stable tetraquark about 20 MeV below the $\bar{B}^*\bar{B}$ threshold. The mass and width of the low-mass $0(1^+)$ $cc\bar{n}\bar{n}$ ($n=u,d$) tetraquark are compatible with the $T_{cc}(3875)^+$ observed by the LHCb Collaboration. The location of the lowest $0(0^+)$ and $0(1^+)$ $bc\bar{n}\bar{n}$ states are found to be close to the $\bar{B}D$ and $\bar{B}^*D$ thresholds, respectively. We hope that the predicted ratios between partial widths of different channels may be helpful to identify compact tetraquark states from future measurements.
\end{abstract}


\maketitle

\section{Introduction}\label{sec1}

In 2021, an exotic narrow state $T_{cc}(3875)^+$ was observed by the LHCb Collaboration in the $D^0D^0\pi^+$ mass spectrum \cite{LHCb:2021vvq}.  This state with width $\Gamma=410\pm165\pm43^{+18}_{-38}$ keV is just $273\pm61\pm5^{+11}_{-14}$ keV below the $D^{*+}D^0$ threshold by using the Breit-Wigner (BW) parameterization. The LHCb also studied the amplitude pole relative to the $D^{*+}D^0$ threshold within a unitarized three-body BW model, which gives a mass $360\pm40^{+4}_{-0}$ keV lower than the $D^{*+}D^0$ threshold and a width $48\pm2^{+0}_{-14}$ keV \cite{LHCb:2021auc}. The minimal quark content of the $T_{cc}(3875)^+$ is $cc\bar{u}\bar{d}$ and its quantum numbers $I(J^P)=0(1^+)$ are favored \cite{LHCb:2021vvq,LHCb:2021auc}. In addition to this exotic state, a search for tetraquark states with the $cc\bar{s}\bar{s}$ configuration was performed by the Belle Collaboration in the invariant mass spectrum of $D_s^+D_s^+(D_s^{*+}D_s^{*+})$ in the $\Upsilon(1S,2S)\to D_s^+D_s^+(D_s^{*+}D_s^{*+})$+anything and $e^+e^-\to D_s^+D_s^+(D_s^{*+}D_s^{*+})$+anything processes \cite{Belle:2021kub}. However, no peaking structures were observed.

The observed $T_{cc}(3875)^+$ is consistent with the expected tetraquark $T_{cc}$ which was widely studied in the literature \cite{Ballot:1983iv,Zouzou:1986qh,Semay:1994ht,Pepin:1996id,Brink:1998as,Ebert:2007rn,Lee:2007tn,Navarra:2007yw,Detmold:2007wk,Lee:2009rt,Yang:2009zzp,Brown:2012tm,Du:2012wp,Hyodo:2012pm,Chen:2013aba,Feng:2013kea,Ikeda:2013vwa,Bicudo:2015vta,Francis:2016hui,Wang:2017dtg,Karliner:2017qjm,Luo:2017eub,Park:2018wjk,Carames:2018tpe,Francis:2018jyb,Junnarkar:2018twb,Deng:2018kly,Agaev:2019kkz,Agaev:2019lwh,Tang:2019nwv,Leskovec:2019ioa,Hernandez:2019eox,Bedolla:2019zwg,Yu:2019sxx,Yang:2019itm,Wallbott:2020jzh,Tan:2020ldi,Lu:2020rog,Yang:2020fou,Braaten:2020nwp,Qin:2020zlg,Cheng:2020nho,Cheng:2020wxa,Gao:2020bvl,Wang:2020jgb,Agaev:2020zag,Hudspith:2020tdf,Mohanta:2020eed}. One may consult Ref. \cite{Liu:2019zoy} for more related discussions. The observation of the exotic $T_{cc}(3875)^+$ triggered further investigations on the $QQ\bar{q}\bar{q}$ ($Q=c,b$ and $q=u,d,s$) tetraquark states. The nature of the doubly charmed $T_{cc}(3875)^+$ has been understood in the $DD^*$ molecule picture \cite{Li:2021zbw,Meng:2021jnw,Chen:2021vhg,Ling:2021bir,Feijoo:2021ppq,Yan:2021wdl,Dai:2021wxi,Wang:2021yld,Xin:2021wcr,Huang:2021urd,Fleming:2021wmk,Ren:2021dsi,Albaladejo:2021vln,Chen:2021tnn,Zhao:2021cvg,Ke:2021rxd,Agaev:2022ast,Cheng:2022qcm,Wang:2022jop,Ortega:2022efc,Lyu:2023xro,Du:2023hlu,Chen:2023fgl,Wang:2023iaz,Dai:2023cyo,Qiu:2023uno,Asanuma:2023atv,Zhai:2023ejo,Guo:2023xyf,Sakai:2023syt} and the compact $cc\bar{u}\bar{d}$ picture \cite{Agaev:2021vur,Wu:2022gie,Liu:2023vrk,He:2023ucd,Meng:2023for,Ma:2023int,Meng:2023jqk,Mutuk:2023oyz} with various methods. The quark-level models \cite{Weng:2021hje,Guo:2021yws,Chen:2021tnn,Jin:2021cxj,Andreev:2021eyj,Deng:2021gnb,Chen:2022ros,Kim:2022mpa,Ortega:2022efc,Wang:2022clw,Meng:2023jqk}, the QCD sum rule \cite{Agaev:2021vur,Ozdem:2021hmk,Albuquerque:2023rrf}, the symmetry-based effective approaches \cite{Chen:2021vhg,Feijoo:2021ppq,Wang:2021yld,Ren:2021dsi,Dai:2021vgf,Du:2021zzh,Ling:2021bir,Yan:2021wdl,Braaten:2022elw,Lin:2022wmj,Wang:2022jop,Dai:2023mxm,Qiu:2023uno,Guo:2023xyf}, the lattice QCD simulations \cite{Padmanath:2022cvl,Meinel:2022lzo,Lyu:2023xro,Asanuma:2023atv,Zhai:2023ejo,Meng:2023bmz}, and so on \cite{Chen:2022asf} have been adopted. Because the mass of $T_{cc}(3875)^+$ is close to the $D^{*+}D^0$ threshold and its decay width is very narrow, the $DD^*$ molecule interpretation has garnered significant attention. However, the possibility of describing this doubly charmed exotic state as a compact tetraquark is not excluded. To understand the nature of $T_{cc}(3875)^+$, its production in various processes is also studied \cite{Hua:2023zpa,Huang:2021urd,Hu:2021gdg,He:2022rta,Abreu:2022lfy,Abreu:2022vmj,Li:2023hpk,Chen:2023xhd,Wang:2023uqd,Hua:2023zpa}.

For studies of $T_{cc}(3875)^+$ in the compact picture, there are theoretical discussions in the diquark-antidiquark configuration \cite{Agaev:2021vur,Wu:2022gie}, the antidiquark-quark-quark configuration \cite{Kim:2022mpa}, and the general four-quark configuration \cite{Chen:2021tnn,Wang:2022clw,Albuquerque:2023rrf,Meng:2023for,Ma:2023int,He:2023ucd,Liu:2023vrk,Mutuk:2023oyz}. The resulting properties are different. If Goldstone boson exchanges are considered in the quark model, deeply bound diquark-antidiquark state \cite{Chen:2021tnn} is possible. In a benchmark test calculation, it is found that there is a tendency for the chiral quark model to overestimate the binding energy \cite{Meng:2023jqk}. The purpose of the present study is to investigate doubly heavy tetraquark states $QQ\bar{q}\bar{q}$ systematically with only one-gluon-exchange interaction. From the results, one discusses whether the observed $T_{cc}(3875)^+$ is a compact state or not and whether stable $QQ\bar{q}\bar{q}$ tetraquarks are possible.

Previously, the study of Ref. \cite{Luo:2017eub} where the masses of $QQ\bar{q}\bar{q}$ tetraquark states are estimated with meson-meson thresholds in a chromomagnetic interaction (CMI) model indicates that several states including the lowest $I(J^P)=0(1^+)$ $cc\bar{u}\bar{d}$ are stable. Later in Ref. \cite{Cheng:2020wxa}, we studied the $QQ\bar{q}\bar{q}$ masses again using a different idea. We adopted the heavy diquark-antiquark symmetry (HDAS) which can relate the high-spin $QQ\bar{q}\bar{q}$ tetraquarks to the $QQq$ baryons. With the mass splittings between $QQ\bar{q}\bar{q}$ states in the CMI model, one gets all the tetraquark masses. The results show that the lowest $I(J^P)=0(1^+)$ $bb\bar{n}\bar{n}$ and the lowest $1/2(1^+)$ $bb\bar{n}\bar{s}$ states are bound, but other states are not stable. To understand the nature of multiquark states, it has been realized that the rearrangement decay properties also play an essential role \cite{Cheng:2019obk,Li:2023aui,Li:2023wxm}. However, only decay patterns were mentioned in Ref. \cite{Luo:2017eub}. To get a better understanding about the doubly heavy tetraquarks, in the present work, we consider both their spectra and their rearrangement decay widths in a simple model.

The decay width of a hadron depends on its mass and the mass in CMI model relies on the estimation method. Comparison of results in different methods and experimental measurements is helpful for us to test theoretical models. Following previous studies of various compact tetraquark states \cite{Wu:2018xdi,Cheng:2020nho,Li:2023wxm}, here we estimate the masses and rearrangement decays for the $QQ\bar{q}\bar{q}$ systems by treating the $X(4140)$ as a reference tetraquark state. We still assume that the decay Hamiltonian is described by a dimensional constant.

This paper is arranged as follows. The formalism to investigate the spectra and decay widths for the doubly heavy tetraquark states in the adopted model will be given in Sec. \ref{sec2}. The numerical results will be given in Sec. \ref{sec3}. The last section presents a brief summary and some discussions.

\section{Formalism }\label{sec2}

\subsection{Mass splitting model}

The interaction force between quark components is mediated by gluons and/or chiral fields in constituent quark models. Usually, the study of a 4-body problem needs the definition of convenient Jacobian coordinates in coordinate or momentum space. Then one gets the spacial wavefunctions and spectrum for the considered tetraquark system by solving the bound state equation after the flavor-spin-color wavefunctions are constructed. Details for such calculations can be found in, e.g., Refs. \cite{Park:2018wjk,Tan:2020ldi,Yang:2009zzp,Lu:2020rog,Wang:2022clw,Noh:2021lqs,Meng:2021yjr}. In the present investigation, our purpose is to check whether one can describe all the compact tetraquark states consistently by focusing on their mass splittings. We adopt the simplified CMI model to estimate tetraquark masses. The symmetry analysis using this model is convenient for us to understand the basic features of the multiquark states easily.

The model Hamiltonian contains the color-magnetic interactions between quark components and it reads
\begin{eqnarray}\label{hamiltonian}
H=\sum_i m_i+H_{\mathrm{CMI}}=\sum_i m_i-\sum_{i<j}C_{ij}\lambda_i\cdot\lambda_j\sigma_i\cdot\sigma_j,
\end{eqnarray}
where the subscript $i$ represents the $i$th constituent quark of the considered state, $m_i$ is its effective mass, and $\lambda_i$ and $\sigma_i$ denote the SU(3) Gell-Mann matrices and SU(2) Pauli matrices for the $i$th quark, respectively. When the quark component is an antiquark, $\lambda_i$ should be replaced with $-\lambda_i^*$. The effective coupling constant $C_{ij}$ describes the strength between the $i$th and $j$th quark components. This Hamiltonian can be understood from that of the quark potential model (QPM) which has the form (see e.g. Ref. \cite{DeRujula:1975qlm})
\begin{eqnarray*}
H_{QPM}=T+V_{CE}+V_{CM}
\end{eqnarray*}
with $T$ being the kinetic term, $V_{CE}$ and $V_{CM}$ being color-electric plus confinement part and color-magnetic part of the potential, respectively. The $V_{CE}$ part is irrelevant with spin-spin interaction of quarks while the $V_{CM}$ part is the contact term described with the delta function. After integrating out the spacial wavefunctions of ground hadrons, one may use the resulting Hamiltonian \eqref{hamiltonian} where $H_{CMI}$ corresponds to $V_{CM}$ to roughly describe the hadron spectra. In the CMI model of multiquark states, only flavor-spin-color wave functions are needed. The
hadron mass formula is expressed with the eigenvalue $E_{CMI}$ of the CMI matrix $\langle H_{CMI}\rangle$, 
\begin{eqnarray}\label{mCMI}
M_1 =\sum_i m_i+E_{\mathrm{CMI}}.
\end{eqnarray}

In principle, the effective parameters ($m_i$ and $C_{ij}$) differ from system to system since the spacial structures affect their values. Without considering such structures explicitly, one cannot determine all of them. In the application of the simplified CMI model, it is usually assumed that the parameters for different states are the same, which obviously gives a worse description of hadron spectrum than the potential quark model. As a result, when a set of parameters is chosen, one probably gets reasonable masses for some hadrons but masses with large errors for other hadrons. If one extracts the involved parameters using a fitting procedure in this crude model, the goodness of fit is low ($\chi^2/d.o.f>10$). Practically, one usually extracts the values of parameters from the masses of part of conventional hadrons by choosing some method. Although the obtained hadron masses in this model probably have large uncertainties, the mass splittings among partner states for which the quark content is the same should be relatively accurate.

Following the procedure used in the literature \cite{Maiani:2004vq,Buccella:2006fn,Kim:2016tys,Luo:2017eub}, we determine the involved parameters $C_{ij}$'s and $m_i$'s. In table \ref{effectiveparameters2}, we show the related hadrons, their CMI expressions, and measured masses \cite{ParticleDataGroup:2022pth}. One extracts the coupling  parameters by comparing the CMI expressions between those in the second and fifth columns. For example, the $C_{c\bar{c}}$ can be determined with $\frac{3}{64}(M_{J/\psi}-M_{\eta_c})$ by using Eq. \eqref{mCMI}. The numerical errors of $C_{ij}$'s are also presented. One obtains them using the error propagation formula. Because the systematic error of the CMI model cannot be estimated and it might be larger than the measurement errors, we do not consider them in the following numerical estimations. We just take $C_{nn}=18.3$ MeV, $C_{ns}=12.1$ MeV, $C_{c\bar{c}}=5.3$ MeV, $C_{b\bar{b}}=2.9$ MeV, $C_{b\bar{n}}=2.1$ MeV, $C_{b\bar{s}}=2.3$ MeV, $C_{b\bar{c}}=3.3$ MeV, $C_{c\bar{s}}=6.7$ MeV, $C_{c\bar{n}}=6.6$ MeV, and $C_{ss}=6.5$ MeV. The parameters $C_{cc}$, $C_{bb}$, and $C_{bc}$ are obtained by using the approximation $\frac{C_{cc}}{C_{c\bar{c}}}=\frac{C_{bb}}{C_{b\bar{b}}}=\frac{C_{bc}}{C_{b\bar{c}}}=\frac{C_{nn}}{C_{n\bar{n}}}\approx\frac23$. The numerical results are  $C_{cc}=3.6$ MeV, $C_{bc}=2.2$ MeV, and $C_{bb}=1.9$ MeV, respectively. The values of parameters we adopt here may be slightly different from those in our previous studies, which is due to the update for the measured hadron masses.
To get the hadron masses with Eq. \eqref{mCMI}, we use the effective quark masses $m_n=361.8$ MeV, $m_s=542.4$ MeV, $m_c=1724.1$ MeV, and $m_b=5054.4$ MeV. They are determined from masses of the baryon states $N$, $\Omega$, $\Sigma_c$, and $\Sigma_b$, respectively. The results are consistent with those in Refs. \cite{Buccella:2006fn,Karliner:2014gca}.

\begin{table}[!htb]
	\caption{Adopted hadrons, related CMI expressions, measured masses \cite{ParticleDataGroup:2022pth}, and the extracted effective coupling parameters.  The values are given in units of MeV. Since the excited $B^*_c$ meson has not been observed in experiments so far, its mass is determined with $m_{B^*_c}-m_{B_c}=70$ MeV \cite{Godfrey:1985xj}.}\setlength{\tabcolsep}{1.3mm}\label{effectiveparameters2}
	\centering
	\begin{tabular}{ccccccc} \hline\hline
		Hadron      &$\langle H_{CMI}\rangle$ & Mass        &Hadron   &$\langle H_{CMI}\rangle$&Mass&$C_{ij}$ \\\hline
		$N$       &$-8C_{nn}$   &       $938.27\pm0.00000029$     &$\Delta$ &$8C_{nn}$   &   $1231.0$         &$C_{nn}=18.30$     \\
		$\pi$     &$-16C_{n\bar{n}}$   & $139\pm0.00018$          &$\rho$   &$\frac{16}{3}C_{n\bar{n}}$  &   $775.26\pm0.23$     &$C_{n\bar{n}}=29.8\pm0.01$     \\
		
		$\Sigma$  &$\frac83C_{nn}-\frac{32}{3}C_{ns}$&$1189.37\pm0.07$&$\Sigma^*$ &$\frac83C_{nn}+\frac{16}{3}C_{ns}$ &$1382.83\pm0.34$ &$C_{ns}=12.09\pm0.03$     \\
		$\eta_c$&$-16C_{c\bar{c}}$&$2983.9\pm0.4$&$J/\psi$&$\frac{16}{3}C_{c\bar{c}}$&$3096.900\pm0.006$&$C_{c\bar{c}}=5.30\pm 0.02$\\
		$\eta_b$&$-16C_{b\bar{b}}$&$9398.70\pm2.0$&$\Upsilon$&$\frac{16}{3}C_{b\bar{b}}$&$9460.30\pm0.26$&$C_{b\bar{b}}=2.89\pm0.09$\\
		
		$B$&$-16C_{b\bar{n}}$&$5279.34\pm0.12$&$B^*$&$\frac{16}{3}C_{b\bar{n}}$&$5324.71\pm0.21$&$C_{b\bar{n}}=2.13\pm0.01$\\
		$B_s$&$-16C_{b\bar{s}}$&$5366..92\pm0.10$&$B^*_s$&$\frac{16}{3}C_{b\bar{s}}$&$5415.4^{+1.8}_{-1.5}$&$C_{b\bar{s}}=2.27\pm0.11$\\
		$B_c$&$-16C_{b\bar{c}}$&$6274.47\pm0.32$&$B^*_c$&$\frac{16}{3}C_{b\bar{c}}$&6344.47&$C_{b\bar{c}}=3.31\pm0.02$\\
		
		$D_s$&$-16C_{c\bar{s}}$&$1968.35\pm0.07$&$D^*_s$&$\frac{16}{3}C_{c\bar{s}}$&$2112.2\pm0.4$&$C_{c\bar{s}}=6.74\pm 0.02$ \\
		$D$&$-16C_{c\bar{n}}$&$1869.66\pm0.05$&$D^*$&$\frac{16}{3}C_{c\bar{n}}$&$2010.26\pm0.05$&$C_{c\bar{n}}=6.591\pm0.003$ \\
		$2\Omega+\Delta-(2\Xi^*+\Xi)$&$8C_{ss}+8C_{nn}$&$198.50\pm0.887$&$(\Delta-N)/2$&$8C_{nn}$&$146.87$&$C_{ss}=6.45\pm0.11$\\
		\hline \hline
	\end{tabular}
\end{table}

\begin{table}[htbp]
	\caption{Comparison for hadron masses measured by experiment (Ex.) and calculated by using Eq. \eqref{mCMI} (Th.). $\Delta M_{\text{Th.}} $ and $\Delta M_{\text{Ex.}}$ denote the calculated and measured mass splittings between hadrons in the 4th and 1st columns, respectively. }\setlength{\tabcolsep}{1.3mm}\label{Dev}
	\centering
	\begin{tabular}{ccccccc} \hline\hline
		Hadron      &Th.         & (Th.-Ex.)  &	Hadron      &Th.         &(Th.-Ex.)&$\Delta M_{\text{Th.}}$($\Delta M_{\text{Ex.}}$)    \\\hline
		$N$         &  939.0        &    1       &$\Delta$      &1231.8         &1 & 292.9(292.7)         \\
		$\pi$       &  246.8        &107         &$\rho$        &882.5          &107&635.7(636.3)\\
		$\Sigma$    &  1185.7       &    -4      &$\Sigma^*$    &1379.3         &-4& 193.6(193.4)          \\
		$\eta_c$    &  3363.4       &    380      &$J/\psi$     &3476.5         &380 &  131.1(131.0)     \\
		$\eta_b$    &  10062.4      &    664      &$\Upsilon$   &10124.3        &664&    61.9(61.6)    \\
		$B$        & 5382.6        &    103      &$B^*$        &5427.4         &103  &   44.8(45.4)   \\
		$B_s$      & 5560.0        &    193      &$B_s^*$      &5609.1         &194 &    49.1(48.5)   \\
		$B_c$      & 6725.7        &    451      &$B_c^*$      &6796.1         &  &      \\
		$D_s$      &  2159.3       &    191      &$D_s^*$      & 2302.2        & 190 & 142.9(143.9)     \\
		$D$        & 1980.3        &    111      &$D^*$        & 2121.1        & 111 & 	140.8(140.6)     \\
		$\Xi$     & 1334.9        &    20      &$\Xi^*$       & 1528.5        & -3   &    193.6(216.9)\\	
		$\Omega$   & 1679.2        &    6.7      &        &  &     &        \\		
		\hline \hline
	\end{tabular}
\end{table}
Using the above parameters, one may get the masses of conventional hadrons with Eq. \eqref{mCMI} again and check the differences between model calculation and experimental measurement. Such results are given in table \ref{Dev}. Obviously, the mass splittings are relatively reasonable, but the hadron masses are usually overestimated. Applied to tetraquark and pentaquark states \cite{Luo:2017eub,Liu:2019zoy,Wu:2018xdi,Li:2018vhp,Wu:2016gas,Zhou:2018bkn,Chen:2016ont,Wu:2016vtq,Wu:2017weo,An:2019idk}, Eq. \eqref{mCMI} also led to overestimated masses. The overestimation means that attractions between quarks are not sufficiently included in the parameters. Noticing the values of parameters, the overestimation should mainly be due to the quark masses. Taking smaller masses is appropriate for the description so that the missed attractions can be compensated. In Ref. \cite{Stancu:2009ka}, it had been questioned whether it is appropriate to apply the quark masses extracted from conventional hadrons to multiquark states. To reduce the uncertainty in multiquark mass estimations, instead of using different values of quark masses, we prefer to use an alternative formula. 

In Ref. \cite{Luo:2017eub}, the masses for the $QQ\bar{q}\bar{q}$ systems were systematically evaluated with a modified formula in the CMI model by introducing a reference system,
\begin{eqnarray}\label{mref}
M_2 =[M_{ref}-(E_{\mathrm{CMI}})_{ref}]+E_{\mathrm{CMI}},
\end{eqnarray}
where $M_{ref}$ and $(E_{\mathrm{CMI}})_{ref}$ denote the threshold and calculated CMI eigenvalue for the reference meson-meson system having the same quark content as the studied state, respectively. Although this equation is just a variant of Eq. \eqref{mCMI}, $M_{ref}$ is the threshold from experiments rather than the calculated one. These two formulas give different results. Therefore, the missed attraction when using Eq. \eqref{mCMI} is partly included in $M_{ref}$. This formula had been employed in Refs. \cite{Leandri:1989su,Silvestre-Brac:1992xsl,Silvestre-Brac:1992kaa}. In fact, Eq. \eqref{mref} is a different model from Eq. \eqref{mCMI}. We will call it threshold scheme. Based on investigations with this method \cite{Luo:2017eub,Wu:2016gas,Chen:2016ont,Wu:2016vtq,Wu:2017weo,Wu:2018xdi,An:2019idk,An:2019idk}, one finds that the obtained masses of multiquark states are generally underestimated. This means that the additional attractions are included when describing multiquark spectra with this scheme. Another problem in this method is the uncertainty of multiquark mass induced by the choice of hadron-hadron threshold. To get a more reasonable mass formula, one needs to phenomenologically fix a reference multiquark state.

In studies of Refs. \cite{Wu:2018xdi,Cheng:2020nho}, we chose the exotic $X(4140)$ as the reference state by assuming it to be the lowest $1^{++}$ $cs\bar{c}\bar{s}$ tetraquark. In the present investigation, we still adopt such an assumption. The considerations for using this assumption are as follows. First, the $X(4140)$ as a $J/\psi \phi$ resonance has been confirmed by different experiments and its quantum numbers were determined to be $J^{PC}=1^{++}$ \cite{ParticleDataGroup:2022pth}. Secondly, the exotic $X(4274)$ observed in the same channel by different experiments \cite{CDF:2011pep,LHCb:2021uow} also has the quantum numbers $J^{PC}=1^{++}$. As partner states containing strange and antistrange quarks, they can be consistently interpreted as compact tetraquark states \cite{Stancu:2009ka,Wu:2016gas}. In addition, we discussed the selection of input state in Ref. \cite{Li:2023wxm} and found that utilizing the reference state $X(4140)$ can provide more reasonable explanations for other observed $cs\bar{c}\bar{s}$ tetraquark states. 

To study various doubly heavy tetraquarks with the above assumption, we will use the following modified mass formula for our discussions,
\begin{eqnarray}\label{mass}
M_3=\tilde{m}+\sum_{ij}n_{ij}\Delta_{ij}+E_{CMI},
\end{eqnarray}
where $\tilde{m}=M_{X(4140)}-(E_{CMI})_{X(4140)}=4231.05$ MeV \cite{Li:2023wxm}, $\Delta_{ij}=m_i-m_j$, and $n_{ij}$ is an integer. The $\Delta_{ij}$ reflects the effective mass gap between the $i$th and $j$th quark components and it is extracted from various conventional hadrons \cite{Wu:2018xdi,Cheng:2020nho}. Because the input parameters in this formula are different from those in the above two formulas, it is actually a third model. We will call it mass splitting scheme. In the present $QQ\bar{q}\bar{q}$ case, the explicit mass formulas for various systems are
\begin{eqnarray}
&&M_{cc\bar{n}\bar{n}}=\tilde{m}+ E_{CMI} -2\Delta_{sn}\label{formula-ccnn},\\
&&M_{cc\bar{n}\bar{s}}=\tilde{m}+ E_{CMI} -\Delta_{sn}\label{formula-ccns},\\
&&M_{cc\bar{s}\bar{s}}=\tilde{m}+ E_{CMI}\label{formula-ccss},\\
&&M_{bc\bar{n}\bar{n}}=\tilde{m}+ E_{CMI} -2\Delta_{sn}+\Delta_{bc}\label{formula-bcnn},\\
&&M_{bc\bar{n}\bar{s}}=\tilde{m}+ E_{CMI} -\Delta_{sn}+\Delta_{bc}\label{formula-bcns},\\
&&M_{bc\bar{s}\bar{s}}=\tilde{m}+ E_{CMI} +\Delta_{bc}\label{formula-bcnn},\\
&&M_{bb\bar{n}\bar{n}}=\tilde{m}+ E_{CMI} -2\Delta_{sn}+2\Delta_{bc}\label{formula-bbnn},\\
&&M_{bb\bar{n}\bar{s}}=\tilde{m}+ E_{CMI} -\Delta_{sn}+2\Delta_{bc}\label{formula-bbns},\\
&&M_{bb\bar{s}\bar{s}}=\tilde{m}+ E_{CMI} +2\Delta_{bc}\label{formula-bbss}.
\end{eqnarray}
The values of the effective mass gaps $\Delta_{bc}$ and $\Delta_{sn}$ have been determined to be $\Delta_{bc}=3340.2$ MeV and $\Delta_{sn}=90.6$ MeV, respectively, in Refs. \cite{Wu:2018xdi,Cheng:2020nho} by analyzing various hadron masses. 

Hadron states definitely have different spacial structures which are affected significantly by the $V_{CE}$ part of $H_{QPM}$. Using the same set of parameters for all the ground hadrons in the CMI model implies that such differences are ignored. Eq. \eqref{mCMI} seems to imply that the spacial structures of overestimated states are enlarged, while Eq. \eqref{mref} seems to imply that the spacial structures of multiquark states are reduced compared to the reference hadron-hadron state. In this sense, Eq. \eqref{mass} can give more reasonable results because both the reference and studied states are compact multiquark states and their spacial structures are comparable. Among the three different estimation methods, the mass order for the same state should be $M_1>M_3>M_2$. Here, one may treat the obtained values with Eq. \eqref{mCMI} as upper limits for the theoretical masses, while the results in the threshold method can just be treated as lower limits for the theoretical masses. Of course, it is possible to narrow the reasonable mass range by using  inequalities \cite{Richard:2019cmi,Cheng:2020wxa} or other theoretical approaches. In the estimation of lower limits for the tetraquark masses with Eq. \eqref{mref}, one does not need effective quark masses but needs reference meson-meson thresholds. The related meson masses have been given in table \ref{effectiveparameters2}. 

Since parameters in the CMI model are extracted from conventional hadrons and the information of spacial wave functions is incorporated in their values, one needs to notice the application of the model. It is generally considered that hadronic molecules are shallow bound states of conventional hadrons bound by meson-exchange forces. Their sizes are usually larger than the conventional hadrons and compact multiquark states. For the interactions between quarks in different hadron components, the adopted parameters are not appropriate. Therefore, we just discuss compact multiquark states with such parameters.

\captionsetup[table]{justification=raggedright}
\begin{table}[htbp]
\caption{All the $color\otimes spin$ wave functions for the $S$-wave $Q_{1}Q_{2}\bar{q}_{3}\bar{q}_{4}$ tetraquark states. The adopted notation is $|(Q_{1}Q_{2})_{color}^{spin}(\bar{q}_{3}\bar{q}_{4})_{color}^{spin}\rangle^{spin}$. The factor $\delta_{ij}$ reflects the constraint from the Pauli principle. If the two heavy quarks are identical, $\delta_{12}=0$. If the two light quarks form a symmetric (antisymmetric) state in flavor space, $\delta_{34}^S=0$ ($\delta_{34}^A=0$). When the factor $\delta_{ij}$ cannot be 0, its value is 1.}
\label{color-spin wave}
\begin{tabular}{c|l|ll}\hline\hline
$J=2$ & $\phi_1\chi_1=|(Q_{1}Q_{2})_{6_c}^{1}(\bar{q}_{3}\bar{q}_{4})_{\bar{6}_c}^{1}\rangle^{2}_{1_c}\delta_{12}\delta_{34}^S$ &$\phi_2\chi_1=|(Q_{1}Q_{2})_{\bar{3}_c}^{1}(\bar{q}_{3}\bar{q}_{4})_{3_c}^{1}\rangle^{2}_{1_c}\delta_{34}^A$\\
\hline
$J=1$&$\phi_1\chi_2=|(Q_{1}Q_{2})_{6_c}^{1}(\bar{q}_{3}\bar{q}_{4})_{\bar{6}_c}^{1}\rangle^{1}_{1_c}\delta_{12}\delta_{34}^S$
&$\phi_2\chi_2=|(Q_{1}Q_{2})_{\bar{3}_c}^{1}(\bar{q}_{3}\bar{q}_{4})_{3_c}^{1}\rangle^{1}_{1_c}\delta_{34}^A$\\
&$\phi_1\chi_4=|(Q_{1}Q_{2})_{6_c}^{1}(\bar{q}_{3}\bar{q}_{4})_{\bar{6}_c}^{0}\rangle^{1}_{1_c}\delta_{12}\delta_{34}^A$
&$\phi_2\chi_4=|(Q_{1}Q_{2})_{\bar{3}_c}^{1}(\bar{q}_{3_c}\bar{q}_{4})_{3}^{0}\rangle^{1}_{1_c}\delta_{34}^S$ \\
&$\phi_1\chi_5=|(Q_{1}Q_{2})_{6_c}^{0}(\bar{q}_{3}\bar{q}_{4})_{\bar{6}_c}^{1}\rangle^{1}_{1_c}\delta_{34}^S$
&$\phi_2\chi_5=|(Q_{1}Q_{2})_{\bar{3}_c}^{0}(\bar{q}_{3_c}\bar{q}_{4})_{3}^{1}\rangle^{1}_{1_c}\delta_{12}\delta_{34}^A$ \\
\hline
$J=0$&$\phi_1\chi_3=|(Q_{1}Q_{2})_{6_c}^{1}(\bar{q}_{3}\bar{q}_{4})_{\bar{6}_c}^{1}\rangle^{0}_{1_c}\delta_{12}\delta_{34}^S$
&$\phi_2\chi_3=|(Q_{1}Q_{2})_{\bar{3}_c}^{1}(\bar{q}_{3}\bar{q}_{4})_{3_c}^{1}\rangle^{0}_{1_c}\delta_{34}^A $\\
&$\phi_1\chi_6=|(Q_{1}Q_{2})_{6_c}^{0}(\bar{q}_{3}\bar{q}_{4})_{\bar{6}_c}^{0}\rangle^{0}_{1_c}\delta_{34}^A$
&$\phi_2\chi_6=|(Q_{1}Q_{2})_{\bar{3}_c}^{0}(\bar{q}_{3}\bar{q}_{4})_{3_c}^{0}\rangle^{0}_{1_c}\delta_{12}\delta_{34}^S$\\ \hline\hline
\end{tabular}
\end{table}
When calculating the CMI matrices for the compact $QQ\bar{q}\bar{q}$ systems, we need their spin ($\chi_{1,\cdots,6}$) and color ($\phi_{1,2}$) wave functions. Here we adopt the diquark-antidiquark type bases which are easy for us to consider the symmetry constraint. Alternatively, one may also adopt the meson-meson type bases and antisymmetrize quark-quark and antiquark-antiquark wave funcions. Since we will diagonalize the obtained matrices, the resulting eigenvalues are independent of choice of bases.  In getting the $color\otimes spin$ wave function bases (table \ref{color-spin wave}), a factor $\delta_{ij}$ is presented so that the symmetry constraint from the Pauli principle can be imposed. In the spin (color) space wave functions, the superscripts (subscripts) denote the spins (color representations) of the diquark and tetraquark states.

\setlength{\tabcolsep}{1.5mm}\begin{table}[htbp]
	\caption{The CMI matrices for the $QQ\bar{q}\bar{q}$ tetraquark states where $n=u$ or $d$.}\scriptsize\label{CMI matrices}
	\begin{tabular}{cccc}\hline\hline
		System & $J^{P}$ &$\langle H_{CMI} \rangle$& Wave function base \\\hline
		$[cc\bar{n}\bar{n}]^{I=1},[bb\bar{n}\bar{n}]^{I=1}, cc\bar{s}\bar{s}, bb\bar{s}\bar{s}$ &$2^+$&$\frac{4}{3}(2\tau+\alpha)$&$(\phi_2\chi_1)^T$\\
		&$1^+$&$\frac{4}{3}(2\tau-\alpha)$&$(\phi_2\chi_2)^T$\\
		&$0^+$&$\begin{pmatrix}
		\frac{8}{3}(\tau-\alpha) &2\sqrt{6}\alpha\\
		&4\tau\\
		\end{pmatrix}$&$(\phi_2\chi_3,\phi_1\chi_6)^T$\\
		$[cc\bar{n}\bar{n}]^{I=0}, [bb\bar{n}\bar{n}]^{I=0}$&$1^+$&$\begin{pmatrix}
		-\frac{8}{3}\eta &2\sqrt{2}\alpha\\
		&\frac{4}{3}\gamma\\
		\end{pmatrix}$&$(\phi_2\chi_4,\phi_1\chi_5)^T$\\
		$[bc\bar{n}\bar{n}]^{I=1},bc\bar{s}\bar{s}$            &$2^+$&$\frac{4}{3}(2\tau+\alpha)$&$(\phi_2\chi_1)^T$\\
		&$1^+$&$\begin{pmatrix}
		\frac{4}{3}(2\tau-\alpha) &-4\mu &-\frac{4\sqrt{2}}{3}\mu \\
		&-\frac{4}{3}\eta&-2\sqrt{2}\tau\\
		&                &-\frac{8}{3}\gamma\\
		\end{pmatrix}$&$(\phi_2\chi_2,\phi_1\chi_4,\phi_2\chi_5)^T$\\
		&$0^+$&$\begin{pmatrix}
		\frac{8}{3}(\tau-\alpha) &2\sqrt{6}\alpha\\
		&4\tau\\
		\end{pmatrix}$&$(\phi_2\chi_3,\phi_1\chi_6)^T$\\
		$[bc\bar{n}\bar{n}]^{I=0}$             &$2^+$&$\frac{2}{3}(-2\tau+5\alpha)$&$(\phi_1\chi_1)^T$\\
		&$1^+$&$\begin{pmatrix}
		-\frac{2}{3}(\tau+5\alpha) &-4\mu &-\frac{10\sqrt{2}}{3}\mu \\
		&\frac{8}{3}\eta&-2\sqrt{2}\alpha\\
		&                &\frac{4}{3}\gamma\\
		\end{pmatrix}$&$(\phi_1\chi_2,\phi_2\chi_4,\phi_1\chi_5)^T$\\
		&$0^+$&$\begin{pmatrix}
		-\frac{4}{3}(\tau+5\alpha) &2\sqrt{6}\alpha\\
		&-8\tau\\
		\end{pmatrix}$&$(\phi_1\chi_3,\phi_2\chi_6)^T$\\
		$cc\bar{n}\bar{s}, bb\bar{n}\bar{s}$                   &$2^+$&$\frac{4}{3}(2\tau+\alpha)$&$(\phi_2\chi_1)^T$\\
		&$1^+$&$\begin{pmatrix}
		\frac{4}{3}(2\tau-\alpha) &\frac{4\sqrt{2}}{3}\beta &4\beta \\
		&\frac{8}{3}\eta&-2\sqrt{2}\alpha\\
		&                &\frac{4}{3}\gamma\\
		\end{pmatrix}$&$(\phi_2\chi_2,\phi_2\chi_4,\phi_1\chi_5)^T$\\
		&$0^+$&$\begin{pmatrix}
		\frac{8}{3}(\tau-\alpha) &2\sqrt{6}\alpha\\
		&4\tau\\
		\end{pmatrix}$&$(\phi_2\chi_3,\phi_1\chi_6)^T$\\
		$bc\bar{n}\bar{s}$                      &$2^+$&$\begin{pmatrix}
		\frac{2}{3}(-2\tau+5\alpha) &2\sqrt{2}\nu\\
		&\frac{4}{3}(2\tau+\alpha)\\
		\end{pmatrix}$&$(\phi_1\chi_1,\phi_2\chi_1)^T$\\
		&$1^+$&$\begin{pmatrix}
		-\frac{2}{3}(2\tau+5\alpha) &2\sqrt{2}\nu &\frac{10\sqrt{2}}{3}\beta &-4\mu &-\frac{10\sqrt{2}}{3}\mu &4\beta \\
		&\frac{4}{3}(2\tau-\alpha)&-4\mu &\frac{4\sqrt{2}}{3}\beta &4\beta &-\frac{4\sqrt{2}}{3}\mu \\
		&                &-\frac{4}{3}\eta &0  &\frac{10}{3}\nu &-2\sqrt{2}\alpha\\
		&                 &                &\frac{8}{3}\eta &-2\sqrt{2}\alpha &\frac{4}{3}\nu\\
		&                 &                &                 &\frac{4}{3}\gamma&0                     \\
		&                  &               &                 &           &-\frac{8}{3}\gamma\\
		\end{pmatrix}$&$\begin{matrix}(\phi_1\chi_2,\phi_2\chi_2,\phi_1\chi_4,\\
		\phi_2\chi_4,\phi_1\chi_5,\phi_2\chi_5)^T\end{matrix}$\\
		&$0^+$&$\begin{pmatrix}
		-\frac{4}{3}(\tau+5\alpha) &4\sqrt{2}\nu &-\frac{10}{\sqrt{3}}\nu &2\sqrt{6}\alpha\\
		&\frac{8}{3}(\tau-\alpha)&2\sqrt{6}\alpha &-\frac{4}{\sqrt{3}}\nu\\
		&2\sqrt{6}\alpha &4\tau &0\\
		&&&-8\tau\\
		\end{pmatrix}$&$\begin{matrix}(\phi_1\chi_3,\phi_2\chi_3,\\
		\phi_1\chi_6,\phi_2\chi_6)^T\end{matrix}$\\\hline\hline
		
	\end{tabular}
\end{table}

With the above constructed wave functions, one obtains the CMI matrices which are listed in table \ref{CMI matrices}. To simplify the expressions, we have defined some variables: $\tau=C_{12}+C_{34}$, $\theta=C_{12}-C_{34}$, $\alpha=C_{13}+C_{14}+C_{23}+C_{24}$, $\beta=C_{13}-C_{14}+C_{23}-C_{24}$, $\mu=C_{13}+C_{14}-C_{23}-C_{24}$, $\nu=C_{13}-C_{14}-C_{23}+C_{24}$, $\gamma=3C_{12}-C_{34}$, $\eta=C_{12}-3C_{34}$. The first six variables are also defined in Ref. \cite{Cheng:2020nho}.

\subsection{Rearrangement decay}

To further understand the nature of multiquark states, a simple decay scheme was adopted to investigate their rearrangement decay properties in our previous works \cite{Cheng:2019obk,Cheng:2020nho,Li:2023aui,Li:2023wxm} where the decay Hamiltonian is described by a constant $H=\mathcal{C}$. The results for the hidden-charm pentaquark states indicate that the estimated decay branching ratios play a role in identifying the structure of the states. Here, we continue to use this simple scheme to study rearrangement decays of the doubly heavy tetraquarks. There are two rearrangement modes,
\begin{eqnarray}
\begin{array}{l}
(Q_1Q_2)(\bar{q_3}\bar{q_4})\to(Q_1\bar{q_3})_{1c}+(Q_2\bar{q_4})_{1c},\\
(Q_1Q_2)(\bar{q_3}\bar{q_4})\to(Q_1\bar{q_4})_{1c}+(Q_2\bar{q_3})_{1c}.
\end{array}
\end{eqnarray}
Their squared amplitude can be expressed as
\begin{eqnarray}
|{\cal M}|^2&=&\mathcal{C}^2|\sum_{ij}x_iy_j|^2,\label{amplitude}
\end{eqnarray}
where $x_i$ satisfying $\sum_{i=1}|x_i|^2=1$ is the element of an eigenvector of the initial state CMI matrix and $y_j$ represents the coefficient when one recouples the final meson-meson state to the $QQ\bar{q}\bar{q}$ base. The width for a rearrangement channel is then given by
\begin{eqnarray}
\Gamma&=&|{\cal M}|^2\frac{\mathbf{|P|}}{8\pi M^2_{QQ\bar{q}\bar{q}}},\label{decay}
\end{eqnarray}
where $M_{QQ\bar{q}\bar{q}}$ represents the mass of the initial tetraquark state and $\mathbf{P}$ is the 3-momentum of a final meson in the rest frame of the tetraquark.

In the adopted decay scheme, the constant $\mathcal{C}$ varies from system to system and it should be extracted from experimentally observed states. In the case of $X(4140)$, one has $\mathcal{C}=7282.15$ MeV \cite{Li:2023wxm}. We use this value to estimate the width of $cc\bar{q}\bar{q}$ states. As a rough estimation, we also use it to evaluate the widths of other $QQ\bar{q}\bar{q}$ states. Once there is an experimental measurement for a system, one should redetermine the value of $\mathcal{C}$ for that system.

\subsection{Effective interactions }

To study the effective interaction between two quark components of a multiquark state, we introduced $K$ factors in Ref. \cite{Li:2018vhp}. The form of this dimensionless measure is given by
\begin{gather}
K_{ij}=\frac{\Delta M}{\Delta C_{ij}},
\end{gather}
where $\Delta C_{ij}$ is the variation of an effective coupling constant and $\Delta M=\Delta \langle H_{\mathrm CMI}\rangle $ is the corresponding variation of the tetraquark mass. If $\Delta C_{ij}$ is small enough, $K_{ij}$ becomes a constant
\begin{gather}
K_{ij}= \lim_{\Delta C_{ij}\to 0}\frac{\Delta M}{\Delta C_{ij}}\to \frac{\partial M}{\partial C_{ij}}.
\end{gather}
Accordingly, the mass formula for a tetraquark state can also be expressed as
\begin{gather}\label{massformulawithK}
M=[M_{ref}-(E_{\mathrm CMI})_{ref}]+\sum_{i<j}K_{ij}C_{ij}.
\end{gather}
The sign of $K_{ij}$ reflects whether the effective CMI between the $i$th and $j$th quark components is attractive ($K_{ij}<0$) or repulsive ($K_{ij}>0$). With the value of a $K_{ij}$, one may also roughly understand the effects on tetraquark mass caused by the uncertainty of the corresponding $C_{ij}$.

\setlength{\tabcolsep}{0.5mm}

\section{Numerical results}\label{sec3}



We obtain the masses and decay widths of all the $QQ\bar{q}\bar{q}$ states and we list them in the following tables \ref{m1}-\ref{decay4}.  In presenting the decay results, we assume that the total width of a tetraquark state is equal to the sum of its two-body rearrangement decay widths, i.e. $\Gamma_{total}\approx\Gamma_{sum}$.

\subsection{$cc\bar{n}\bar{n}$, $cc\bar{n}\bar{s}$, and $cc\bar{s}\bar{s}$ systems }

\begin{figure}[htbp]
	\centering
	\begin{minipage}[b]{0.32\textwidth}
		\centering
		\includegraphics[width=1\textwidth]{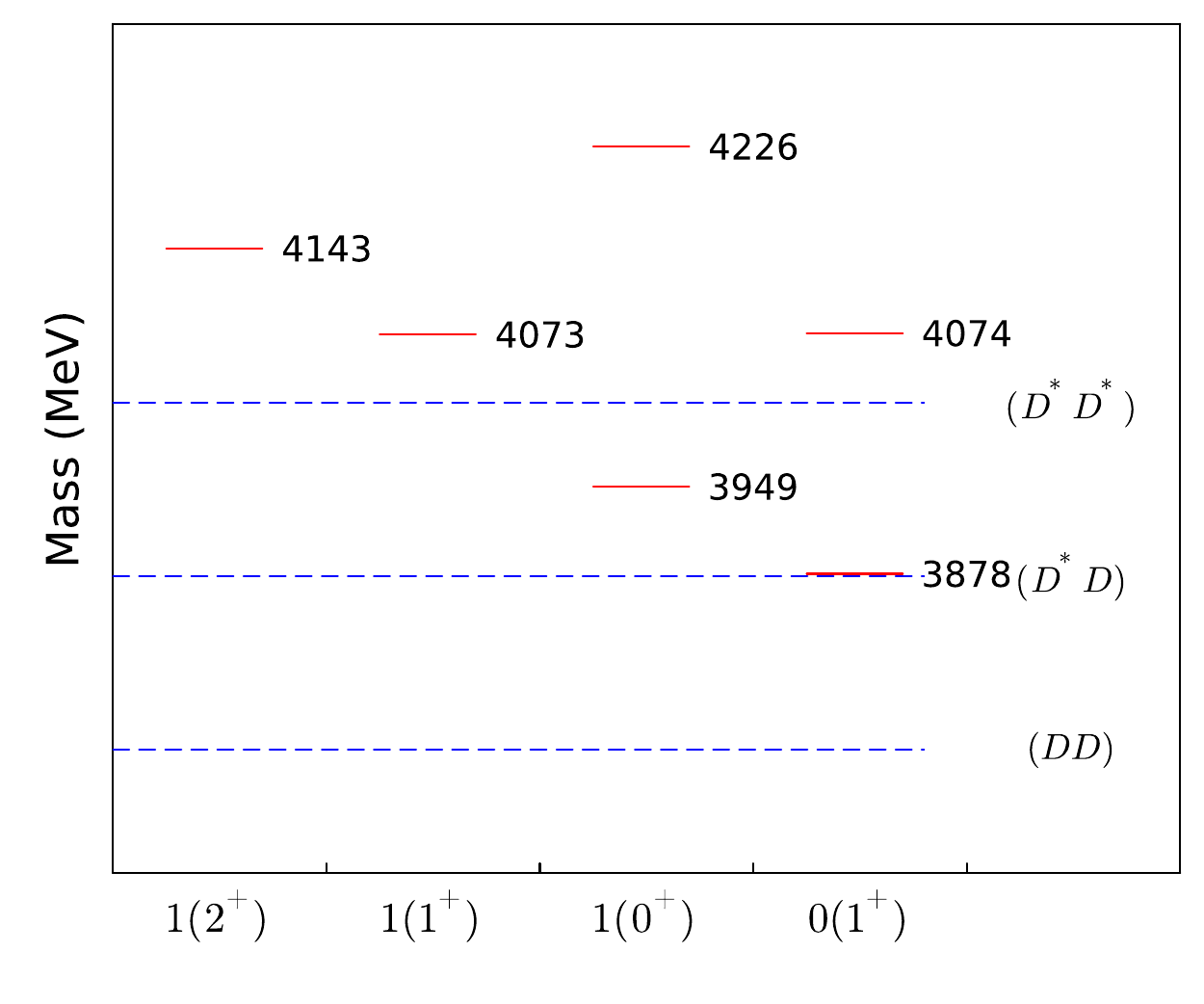}
		\subcaption{$cc\bar{n}\bar{n}$}
	\end{minipage}
	\begin{minipage}[b]{0.32\textwidth}
		\centering
		\includegraphics[width=1\textwidth]{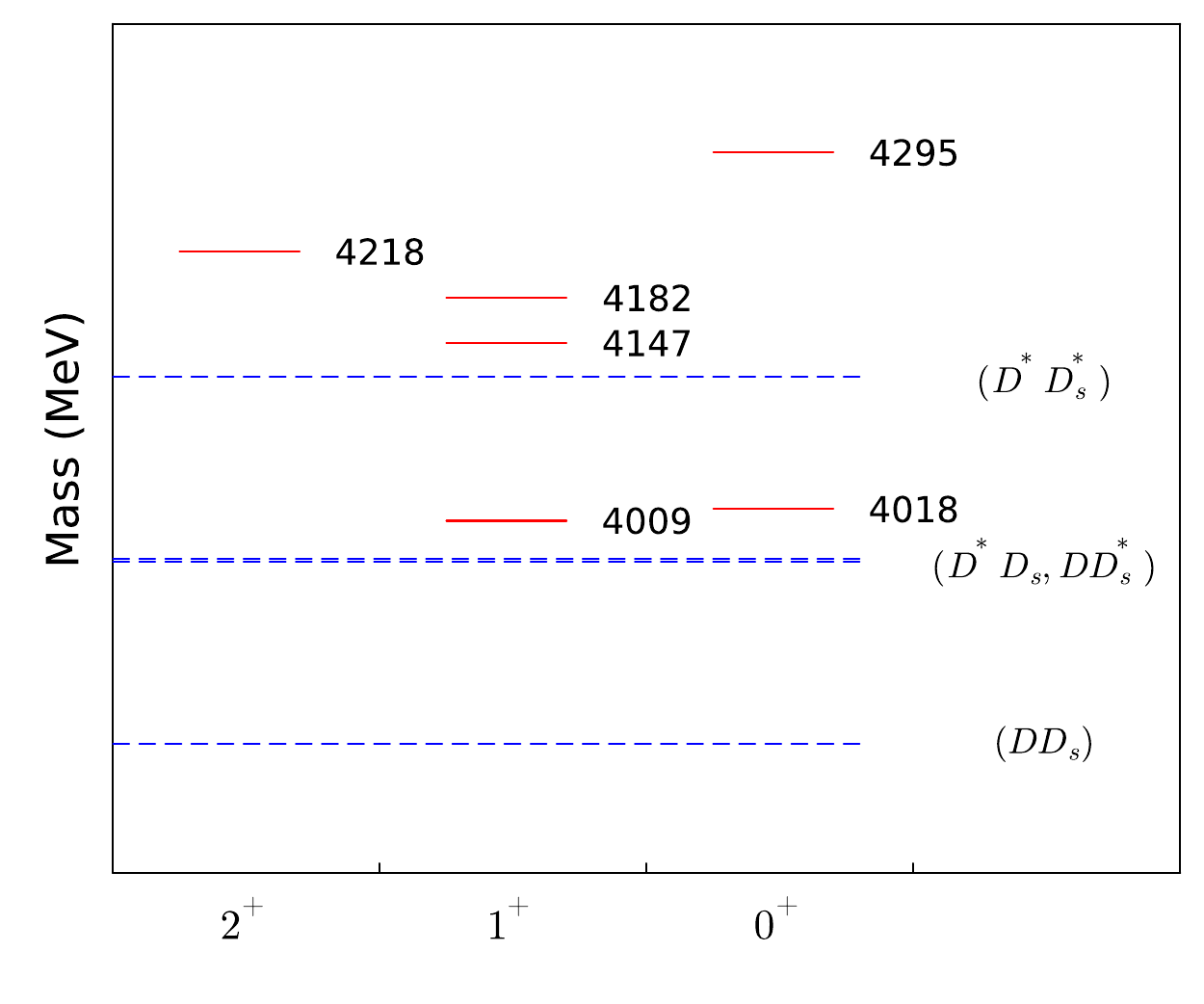}
		\subcaption{$cc\bar{n}\bar{s}$}
	\end{minipage}
	\begin{minipage}[b]{0.32\textwidth}
		\centering
		\includegraphics[width=1\textwidth]{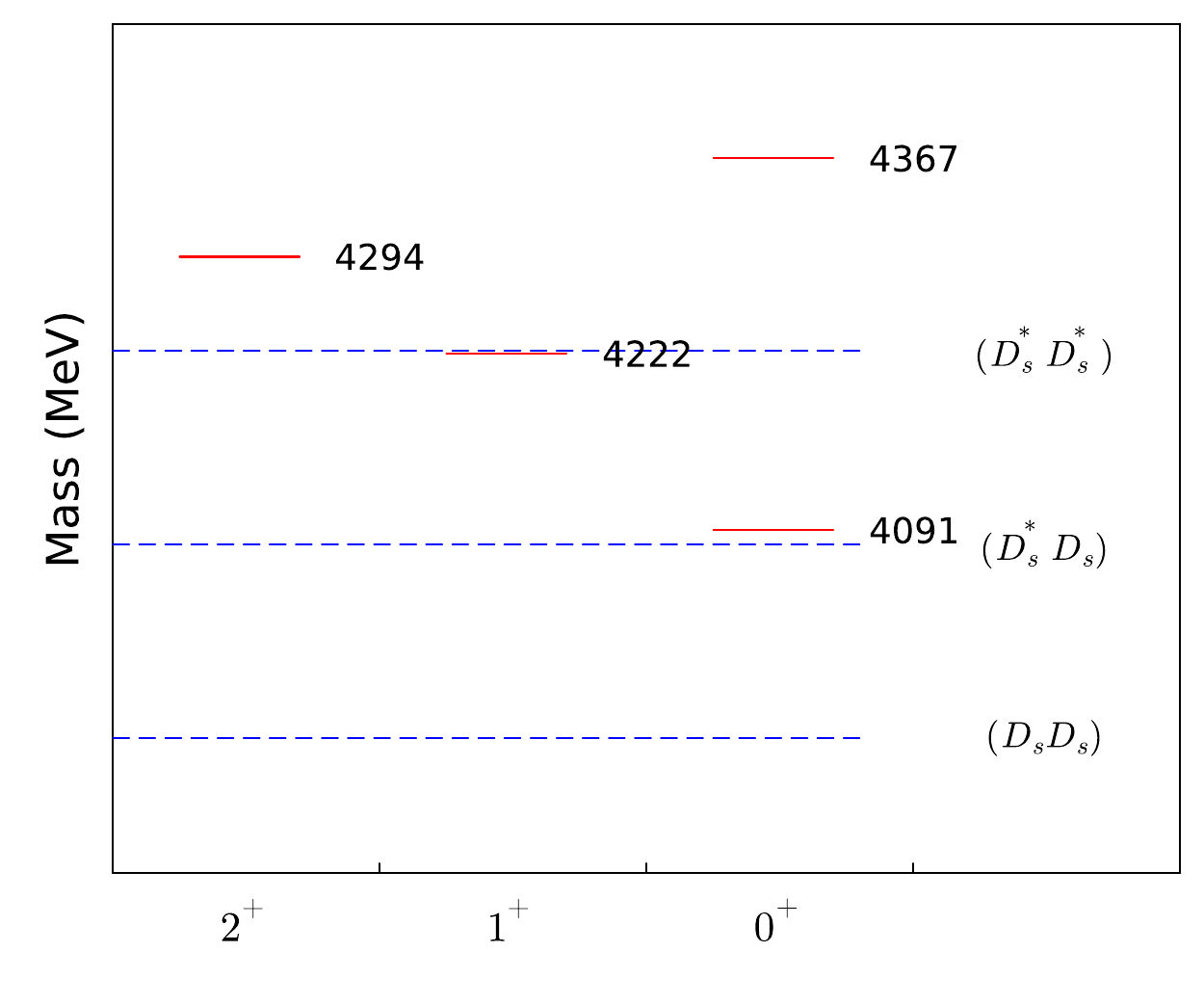}
		\subcaption{$cc\bar{s}\bar{s}$}
	\end{minipage}
	\caption{Relative positions for the considered tetraquark states. The red solid lines represent the estimated masses with the mass splitting scheme. The blue dashed lines indicate various meson-meson thresholds.\label{ccqq-picture}}
\end{figure}

The spectrum for tetraquark states in these systems are presented in table \ref{m1} and their relative positions are displayed in Fig. \ref{ccqq-picture}. With the mass splitting scheme, we obtain higher results than Ref. \cite{Luo:2017eub}, which makes the previous stable states decay. Noticing that the tetraquark masses with the threshold scheme seem to be underestimated, we regard the results with the mass splitting scheme as more reasonable ones.

\begin{table}[!h]\centering
	\caption{Mass spectrum for the $cc\bar{q}\bar{q}$ states in units of MeV. The last three columns list the results obtained with Eqs. \eqref{mass}, \eqref{mref}, and \eqref{mCMI}, respectively. The thresholds of $DD$, $DD_s$, and $D_sD_s$ are used in getting the lower limits for the tetraquark masses in the cases of $cc\bar{n}\bar{n}$, $cc\bar{n}\bar{s}$, and $cc\bar{s}\bar{s}$, respectively.  The upper limits are obtained using Eq\eqref{mCMI}, where the effective quark masses  are $m_n=361.8$ MeV, $m_s=542.4$ MeV, $m_c=1724.1$ MeV, and $m_b=5054.4$ MeV, respectively. We take values in the mass splitting scheme (fourth column) in this article.}\scriptsize\label{m1}
	\begin{tabular}{c|cccccc}\hline
		\hline\multicolumn{6}{c}{$cc\bar{n}\bar{n}$ system} \\\hline\hline
		$I(J^{P})$ & $\langle H_{CMI} \rangle$ & $E_{CMI}$ &Mass&Lower limits&Upper limits\\\hline
		$1(2^{+})$ &$\left(\begin{array}{c}93.3\end{array}\right)$&$\left(\begin{array}{c}93.3\end{array}\right)$&$\left(\begin{array}{c}4143.2\end{array}\right)$&$\left(\begin{array}{c}4034.1\end{array}\right)$&$\left(\begin{array}{c}4265.1\end{array}\right)$\\
		$1(1^{+})$ &$\left(\begin{array}{c}22.9\end{array}\right)$&$\left(\begin{array}{c}22.9\end{array}\right)$&$\left(\begin{array}{c}4072.8\end{array}\right)$&$\left(\begin{array}{c}3963.7\end{array}\right)$&$\left(\begin{array}{c}4194.7\end{array}\right)$\\
		$1(0^{+})$ &$\left(\begin{array}{cc}-12.3&129.3\\129.3&87.2\end{array}\right)$&$\left(\begin{array}{c}176.0\\-101.1\end{array}\right)$&$\left(\begin{array}{c}4225.9\\3948.8\end{array}\right)$&$\left(\begin{array}{c}4116.8\\3839.7\end{array}\right)$&$\left(\begin{array}{c}4347.8\\4070.7\end{array}\right)$\\
		$0(1^{+})$ &$\left(\begin{array}{cc}-137.1&-74.7\\-74.7&-10.4\end{array}\right)$&$\left(\begin{array}{c}24.2\\-171.6\end{array}\right)$&$\left(\begin{array}{c}4074.0\\3878.2\end{array}\right)$&$\left(\begin{array}{c}3965.0\\3769.2\end{array}\right)$&$\left(\begin{array}{c}4196.0\\4000.2\end{array}\right)$\\
		\hline\multicolumn{6}{c}{$cc\bar{n}\bar{s}$ system} \\\hline\hline
		$J^{P}$ & $\langle H_{CMI} \rangle$ & $E_{CMI}$ &Mass&Lower limits&Upper limits\\\hline
		$2^{+}$ &$\left(\begin{array}{c}77.1\end{array}\right)$&$\left(\begin{array}{c}77.1\end{array}\right)$&$\left(\begin{array}{c}4217.5\end{array}\right)$&$\left(\begin{array}{c}4123.0\end{array}\right)$&$\left(\begin{array}{c}4429.5\end{array}\right)$\\
		$1^{+}$ &$\left(\begin{array}{ccc}6.1&-0.4&-0.8\\-0.4&-87.5&-75.2\\-0.8&-75.2&-2.1\end{array}\right)$&$\left(\begin{array}{c}41.7\\6.1\\-131.3\end{array}\right)$&$\left(\begin{array}{c}4182.2\\4146.6\\4009.2\end{array}\right)$&$\left(\begin{array}{c}4087.6\\4052.0\\3914.6\end{array}\right)$&$\left(\begin{array}{c}4394.1\\4358.5\\4221.1\end{array}\right)$\\
		$0^{+}$ &$\left(\begin{array}{cc}-29.3&130.3\\130.3&62.4\end{array}\right)$&$\left(\begin{array}{c}154.7\\-121.6\end{array}\right)$&$\left(\begin{array}{c}4295.1\\4018.8\end{array}\right)$&$\left(\begin{array}{c}4200.6\\3924.3\end{array}\right)$&$\left(\begin{array}{c}4507.1\\4230.8\end{array}\right)$\\
		\hline\multicolumn{6}{c}{$cc\bar{s}\bar{s}$ system} \\\hline\hline
		$J^{P}$ & $\langle H_{CMI} \rangle$ & $E_{CMI}$  &Mass&Lower limits&Upper limits\\\hline
		$2^{+}$ &$\left(\begin{array}{c}62.4\end{array}\right)$&$\left(\begin{array}{c}62.4\end{array}\right)$&$\left(\begin{array}{c}4293.5\end{array}\right)$&$\left(\begin{array}{c}4213.4\end{array}\right)$&$\left(\begin{array}{c}4595.4\end{array}\right)$\\
		$1^{+}$ &$\left(\begin{array}{c}-9.1\end{array}\right)$&$\left(\begin{array}{c}-9.1\end{array}\right)$&$\left(\begin{array}{c}4222.0\end{array}\right)$&$\left(\begin{array}{c}4141.9\end{array}\right)$&$\left(\begin{array}{c}4523.9\end{array}\right)$\\
		$0^{+}$ &$\left(\begin{array}{cc}-44.8&131.3\\131.3&40.0\end{array}\right)$&$\left(\begin{array}{c}135.6\\-140.4\end{array}\right)$&$\left(\begin{array}{c}4366.6\\4090.7\end{array}\right)$&$\left(\begin{array}{c}4286.6\\4010.6\end{array}\right)$&$\left(\begin{array}{c}4668.6\\4392.6\end{array}\right)$\\
		\hline
	\end{tabular}
\end{table}

In a recent work \cite{Kim:2022mpa}, the authors studied the doubly heavy tetraquarks with a chiral-diquark potential model. Their results show that the lowest $1S(1^+)$ $cc\bar{u}\bar{d}$ tetraquark state is around $3961$ MeV (86 MeV above the $D^0D^{*+}$ threshold) and no stable $T_{cc}$ tetraquark state is observed. This conclusion is similar to those drawn in other quark-level model calculations \cite{Wang:2022clw,Cheng:2020wxa,Braaten:2020nwp,Lu:2020rog,Park:2018wjk,Eichten:2017ffp}. Prior to the LHCb measurement, Karliner and Rosner \cite{Karliner:2017qjm} obtained a $T_{cc}$ mass about 7 MeV above the $D^0D^{*+}$ threshold. If one increases the $cc$ binding energy by 6 MeV so that the predicted mass of $\Xi_{cc}$ agrees well with the measured value, the lowest $cc\bar{u}\bar{d}$ with $I(J^P)=0(1^+)$ would be located at 3876 MeV, only 1 MeV above the $D^0D^{*+}$ threshold. In their adopted model, the results indicate that this compact tetraquark is a near-threshold state. In the investigation with the heavy diquark-antiquark symmetry \cite{Cheng:2020wxa}, the lowest $cc\bar{u}\bar{d}$ is $\sim$60 MeV above the $DD^*$ threshold. It is difficult to interpret the LHCb $T_{cc}$ as a compact tetraquark if the estimation is reliable. According to present results in the mass splitting scheme, the lowest $cc\bar{n}\bar{n}$ state ($M=3878.2$ MeV) is very close to ($\sim$3.1 MeV above) the $D^0D^{*+}$ threshold. Based solely on the mass spectrum, 
 our result is compatible with the quark model studies of Refs. \cite{Karliner:2017qjm,Liu:2023vrk,Meng:2023for,Meng:2023jqk} and the LHCb measurements \cite{LHCb:2021vvq,LHCb:2021auc}.

\setlength{\tabcolsep}{0.1mm}\begin{table}[htbp]\caption{$K_{ij}$'s for the $cc\bar{n}\bar{n}$, $cc\bar{n}\bar{s}$, $cc\bar{s}\bar{s}$, $bb\bar{n}\bar{n}$, $bb\bar{n}\bar{s}$, and $bb\bar{s}\bar{s}$ states. The orders of states are consistent with tables \ref{m1} and \ref{m2}.}\label{Kij-ccqq}\scriptsize\centering
	\begin{tabular}{cccccccccc}\hline\hline
		&\multicolumn{4}{c}{$cc\bar{n}\bar{n}$}&&\multicolumn{4}{c}{$bb\bar{n}\bar{n}$}\\
		\hline
		\hline
		$I(J^{P})$&$K_{cc}$&$K_{c\bar{n}}$&$K_{nn}$&&$I(J^{P})$&$K_{bb}$&$K_{b\bar{n}}$&$K_{nn}$&\\
		$1(2^{+})$&$\left[\begin{array}{c}2.7\end{array}\right]$&$\left[\begin{array}{c}5.3\end{array}\right]$&$\left[\begin{array}{c}2.7\end{array}\right]$&&$1(2^{+})$&$\left[\begin{array}{c}2.7\end{array}\right]$&$\left[\begin{array}{c}5.3\end{array}\right]$&$\left[\begin{array}{c}2.7\end{array}\right]$\\
		$1(1^{+})$&$\left[\begin{array}{c}2.7\end{array}\right]$&$\left[\begin{array}{c}-5.3\end{array}\right]$&$\left[\begin{array}{c}2.7\end{array}\right]$&&$1(1^{+})$&$\left[\begin{array}{c}2.7\end{array}\right]$&$\left[\begin{array}{c}-5.3\end{array}\right]$&$\left[\begin{array}{c}2.7\end{array}\right]$\\
		$1(0^{+})$&$\left[\begin{array}{c}3.6\\3.1\end{array}\right]$&$\left[\begin{array}{c}14.9\\-25.5\end{array}\right]$&$\left[\begin{array}{c}3.6\\3.1\end{array}\right]$&&$1(0^{+})$&$\left[\begin{array}{c}3.7\\3.0\end{array}\right]$&$\left[\begin{array}{c}14.2\\-24.9\end{array}\right]$&$\left[\begin{array}{c}3.7\\3.0\end{array}\right]$\\
		$0(1^{+})$&$\left[\begin{array}{c}3.8\\2.9\end{array}\right]$&$\left[\begin{array}{c}8.6\\-8.6\end{array}\right]$&$\left[\begin{array}{c}-2.5\\-6.8\end{array}\right]$&&$0(1^{+})$&$\left[\begin{array}{c}4.0\\2.7\end{array}\right]$&$\left[\begin{array}{c}4.0\\-4.0\end{array}\right]$&$\left[\begin{array}{c}-1.6\\-7.8\end{array}\right]$\\
		\hline
		&\multicolumn{4}{c}{$cc\bar{n}\bar{s}$}&&\multicolumn{4}{c}{$bb\bar{n}\bar{s}$}\\
		\hline
		$J^{P}$&$K_{cc}$&$K_{c\bar{n}}$&$K_{c\bar{s}}$&$K_{ns}$&$J^{P}$&$K_{bb}$&$K_{b\bar{n}}$&$K_{b\bar{s}}$&$K_{ns}$\\
		$2^{+}$&$\left[\begin{array}{c}2.7\end{array}\right]$&$\left[\begin{array}{c}5.3\end{array}\right]$&$\left[\begin{array}{c}5.3\end{array}\right]$&$\left[\begin{array}{c}2.7\end{array}\right]$&$2^{+}$&$\left[\begin{array}{c}2.7\end{array}\right]$&$\left[\begin{array}{c}2.7\end{array}\right]$&$\left[\begin{array}{c}2.7\end{array}\right]$&$\left[\begin{array}{c}2.7\end{array}\right]$\\
		$1^{+}$&$\left[\begin{array}{c}3.7\\2.7\\3.0\end{array}\right]$&$\left[\begin{array}{c}4.8\\-2.6\\-4.8\end{array}\right]$&$\left[\begin{array}{c}5.1\\-2.7\\-5.0\end{array}\right]$&$\left[\begin{array}{c}-3.0\\2.7\\-6.3\end{array}\right]$&$1^{+}$&$\left[\begin{array}{c}2.7\\3.9\\2.8\end{array}\right]$&$\left[\begin{array}{c}-3.4\\3.6\\-2.8\end{array}\right]$&$\left[\begin{array}{c}-1.9\\2.2\\-3.0\end{array}\right]$&$\left[\begin{array}{c}2.7\\-1.8\\-7.5\end{array}\right]$\\
		$0^{+}$&$\left[\begin{array}{c}3.6\\3.1\end{array}\right]$&$\left[\begin{array}{c}7.5\\-12.8\end{array}\right]$&$\left[\begin{array}{c}7.5\\-12.8\end{array}\right]$&$\left[\begin{array}{c}3.6\\3.1\end{array}\right]$&$0^{+}$&$\left[\begin{array}{c}3.6\\3.0\end{array}\right]$&$\left[\begin{array}{c}7.3\\-12.6\end{array}\right]$&$\left[\begin{array}{c}7.3\\-12.6\end{array}\right]$&$\left[\begin{array}{c}3.6\\3.0\end{array}\right]$\\
		\hline
		&\multicolumn{4}{c}{$cc\bar{s}\bar{s}$}&&\multicolumn{4}{c}{$bb\bar{s}\bar{s}$}\\
		\hline
		$J^{P}$&$K_{cc}$&$K_{c\bar{s}}$&$K_{ss}$&&$J^{P}$&$K_{bb}$&$K_{b\bar{s}}$&$K_{ss}$&\\
		$2^{+}$&$\left[\begin{array}{c}2.7\end{array}\right]$&$\left[\begin{array}{c}5.3\end{array}\right]$&$\left[\begin{array}{c}2.7\end{array}\right]$&&$2^{+}$&$\left[\begin{array}{c}2.7\end{array}\right]$&$\left[\begin{array}{c}5.3\end{array}\right]$&$\left[\begin{array}{c}2.7\end{array}\right]$\\
		$1^{+}$&$\left[\begin{array}{c}2.7\end{array}\right]$&$\left[\begin{array}{c}-5.3\end{array}\right]$&$\left[\begin{array}{c}2.7\end{array}\right]$&&$1^{+}$&$\left[\begin{array}{c}2.7\end{array}\right]$&$\left[\begin{array}{c}-5.3\end{array}\right]$&$\left[\begin{array}{c}2.7\end{array}\right]$\\
		$0^{+}$&$\left[\begin{array}{c}3.5\\3.1\end{array}\right]$&$\left[\begin{array}{c}15.0\\-25.6\end{array}\right]$&$\left[\begin{array}{c}3.5\\3.1\end{array}\right]$&&$0^{+}$&$\left[\begin{array}{c}3.6\\3.1\end{array}\right]$&$\left[\begin{array}{c}14.8\\-25.5\end{array}\right]$&$\left[\begin{array}{c}3.6\\3.1\end{array}\right]$\\
		\hline\hline
	\end{tabular}
\end{table}

Table \ref{Kij-ccqq} lists the results about the $K$ factors. From their signs and values and Eq. \eqref{massformulawithK}, the effective attraction between the two light quarks and that between heavy and light quarks are important in lowering the mass of the ground $cc\bar{u}\bar{d}$. The effects on uncertainty of CMI from the uncertainty of $C_{nn}$ and $C_{c\bar{n}}$ may also be understood.

\setlength{\tabcolsep}{0.5mm}\begin{table}[!h]\centering\tiny
	\caption{Rearrangement decays for the $cc\bar{n}\bar{n}$, $cc\bar{n}\bar{s}$, $cc\bar{s}\bar{s}$, $bb\bar{n}\bar{n}$, $bb\bar{n}\bar{s}$, and $bb\bar{s}\bar{s}$ cases. The two numbers in the parenthesis for a decay channel mean dimensionless $100|\mathcal{M}|^2/{\mathcal C}^2$ and dimensional partial width, respectively. The masses and widths are presented in units of MeV.}\label{decay1}
	\begin{tabular}{cccccc|cccccc}\hline\hline
		&&\multicolumn{3}{c}{$cc\bar{n}\bar{n}$}&&&&\multicolumn{3}{c}{$bb\bar{n}\bar{n}$}&\\
		\hline
		$I(J^{P})$&Mass&\multicolumn{3}{c}{Channels}&$\Gamma$&$I(J^{P})$&Mass&\multicolumn{3}{c}{Channels}&$\Gamma$\\
		\hline
		&&$D^{*} D^*$&&&&&&$\bar{B}^{*} \bar{B}^*$&&&\\
		$1(2^{+})$&$\left[\begin{array}{c}4143.2\end{array}\right]$&$\left[\begin{array}{c}(33.3, 20.8)\end{array}\right]$&&&$\left[\begin{array}{c}20.8\end{array}\right]$&$1(2^{+})$&$\left[\begin{array}{c}10795.3\end{array}\right]$&$\left[\begin{array}{c}(33.3, 5.3)\end{array}\right]$&&&$\left[\begin{array}{c}5.3\end{array}\right]$\\
		&&$D^*D $&&&&&&$\bar{B}^*\bar{B} $&&&\\
		$1(1^{+})$&$\left[\begin{array}{c}4072.8\end{array}\right]$&$\left[\begin{array}{c}(16.7, 53.0)\end{array}\right]$&&&$\left[\begin{array}{c}53.0\end{array}\right]$&$1(1^{+})$&$\left[\begin{array}{c}10772.9\end{array}\right]$&$\left[\begin{array}{c}(16.7, 11.5)\end{array}\right]$&&&$\left[\begin{array}{c}11.5\end{array}\right]$\\
		&&$D^*D^* $&$DD$&&&&&$\bar{B}^*\bar{B}^*$&$\bar{B}\bar{B}$&&\\
		$1(0^{+})$&$\left[\begin{array}{c}4225.9\\3948.8\end{array}\right]$&$\left[\begin{array}{c}(55.7, 43.2)\\(2.6, -)\end{array}\right]$&$\left[\begin{array}{c}(0.3, 0.3)\\(41.4, 35.9)\end{array}\right]$&&$\left[\begin{array}{c}43.5\\35.9\end{array}\right]$&$1(0^{+})$&$\left[\begin{array}{c}10834.4\\10738.4\end{array}\right]$&$\left[\begin{array}{c}(57.4, 10.3)\\(0.9, 0.1)\end{array}\right]$&$\left[\begin{array}{c}(1.2, 0.3)\\(40.5, 7.2)\end{array}\right]$&&$\left[\begin{array}{c}10.5\\7.4\end{array}\right]$\\
		&&$D^*D^* $&$D^*D$&&&&&$\bar{B}^*\bar{B}^*$&$\bar{B}^*\bar{B}$&&\\
		$0(1^{+})$&$\left[\begin{array}{c}4074.0\\3878.2\end{array}\right]$&$\left[\begin{array}{c}(48.4, 20.9)\\(1.6, -)\end{array}\right]$&$\left[\begin{array}{c}(6.2, 19.8)\\(18.8, 7.2)\end{array}\right]$&&$\left[\begin{array}{c}40.7\\7.2\end{array}\right]$&$0(1^{+})$&$\left[\begin{array}{c}10717.8\\10584.5\end{array}\right]$&$\left[\begin{array}{c}(41.2, 4.6)\\(8.8, -)\end{array}\right]$&$\left[\begin{array}{c}(12.2, 7.0)\\(12.8, -)\end{array}\right]$&&$\left[\begin{array}{c}11.6\\0\end{array}\right]$\\
		\hline
		&&\multicolumn{3}{c}{$cc\bar{n}\bar{s}$}&&&&\multicolumn{3}{c}{$bb\bar{n}\bar{s}$}&\\
		\hline
		&&$D^*D_s^{*}$&&&&&&$\bar{B}^*\bar{B}_s^{*}$&&&\\
		$2^{+}$&$\left[\begin{array}{c}4217.5\end{array}\right]$&$\left[\begin{array}{c}(33.3, 35.5)\end{array}\right]$&&&$\left[\begin{array}{c}35.5\end{array}\right]$&$2^{+}$&$\left[\begin{array}{c}10869.9\end{array}\right]$&$\left[\begin{array}{c}(33.3, 10.0)\end{array}\right]$&&&$\left[\begin{array}{c}10.0\end{array}\right]$\\
		&&$D^*D_s^* $&$D^*D_s$&$DD_s^*$&&&&$\bar{B}^*\bar{B}_s^* $&$\bar{B}^*\bar{B}_s$&$\bar{B}\bar{B}^*_s$&\\
		$1^{+}$&$\left[\begin{array}{c}4182.2\\4146.6\\4009.2 \end{array}\right]$&$\left[\begin{array}{c}(49.6, 42.7)\\(0.0, 0.0)\\(0.4, -) \end{array}\right]$&$\left[\begin{array}{c}(4.1, 6.4)\\(16.7, 24.1)\\(20.9, 13.9) \end{array}\right]$&$\left[\begin{array}{c}(4.5, 7.0)\\(16.6, 23.7)\\(20.5, 13.1) \end{array}\right]$&$\left[\begin{array}{c}56.1\\47.8\\27.0\end{array}\right]$&$1^{+}$&$\left[\begin{array}{c}10846.5\\10819.1\\10722.2 \end{array}\right]$&$\left[\begin{array}{c}(0.1, 0.0)\\(44.2, 10.4)\\(5.7, -) \end{array}\right]$&$\left[\begin{array}{c}(15.1, 4.9)\\(11.3, 3.4)\\(15.3, 2.3) \end{array}\right]$&$\left[\begin{array}{c}(18.2, 5.9)\\(8.8, 2.6)\\(14.7, 2.1) \end{array}\right]$&$\left[\begin{array}{c}10.9\\16.3\\4.3\end{array}\right]$\\
		&&$D^*D^*_s $&$DD_s$&&&&&$\bar{B}^*\bar{B}^*_s$&$\bar{B}\bar{B}_s$&&\\
		$0^{+}$&$\left[\begin{array}{c}4295.1\\4018.8\end{array}\right]$&$\left[\begin{array}{c}(55.3, 76.7)\\(3.0, -)\end{array}\right]$&$\left[\begin{array}{c}(0.2, 0.4)\\(41.5, 65.0)\end{array}\right]$&&$\left[\begin{array}{c}77.1\\65.0\end{array}\right]$&$0^{+}$&$\left[\begin{array}{c}10903.8\\10807.8\end{array}\right]$&$\left[\begin{array}{c}(56.7, 18.9)\\(1.7, 0.4)\end{array}\right]$&$\left[\begin{array}{c}(0.7, 0.3)\\(41.0, 13.8)\end{array}\right]$&&$\left[\begin{array}{c}19.2\\14.2\end{array}\right]$\\
		\hline
		&&\multicolumn{3}{c}{$cc\bar{s}\bar{s}$}&&&&\multicolumn{3}{c}{$bb\bar{s}\bar{s}$}&\\
		\hline
		&&$D_s^*D_s^{*}$&&&&&&$\bar{B}_s^*\bar{B}_s^{*}$&&&\\
		$2^{+}$&$\left[\begin{array}{c}4293.5\end{array}\right]$&$\left[\begin{array}{c}(33.3, 14.6)\end{array}\right]$&&&$\left[\begin{array}{c}14.6\end{array}\right]$&$2^{+}$&$\left[\begin{array}{c}10946.1\end{array}\right]$&$\left[\begin{array}{c}(33.3, 4.7)\end{array}\right]$&&&$\left[\begin{array}{c}4.7\end{array}\right]$\\
		&&$D_s^*D_s$&&&&&&$\bar{B}_s^*\bar{B}_s$&&&\\
		$1^{+}$&$\left[\begin{array}{c}4222.0\end{array}\right]$&$\left[\begin{array}{c}(16.7, 42.7)\end{array}\right]$&&&$\left[\begin{array}{c}42.7\end{array}\right]$&$1^{+}$&$\left[\begin{array}{c}10921.6\end{array}\right]$&$\left[\begin{array}{c}(16.7, 10.3)\end{array}\right]$&&&$\left[\begin{array}{c}10.3\end{array}\right]$\\
		&&$D^*_sD^*_s $&$D_sD_s$&&&&&$\bar{B}^*_s\bar{B}^*_s$&$\bar{B}_s\bar{B}_s$&&\\
		$0^{+}$&$\left[\begin{array}{c}4366.6\\4090.7\end{array}\right]$&$\left[\begin{array}{c}(55.0, 33.6)\\(3.3, -)\end{array}\right]$&$\left[\begin{array}{c}(0.1, 0.1)\\(41.5, 29.1)\end{array}\right]$&&$\left[\begin{array}{c}33.8\\29.1\end{array}\right]$&$0^{+}$&$\left[\begin{array}{c}10975.7\\10878.7\end{array}\right]$&$\left[\begin{array}{c}(55.8, 8.7)\\(2.5, 0.2)\end{array}\right]$&$\left[\begin{array}{c}(0.3, 0.1)\\(41.3, 6.5)\end{array}\right]$&&$\left[\begin{array}{c}8.8\\6.8\end{array}\right]$\\
		\hline\hline
	\end{tabular}
\end{table}

The rearrangement decay properties for the $cc\bar{n}\bar{n}$ states are given in table \ref{decay1}. It is shown that the decay width of the lowest $0(1^+)$ state is about $7.2$ MeV and its main decay channel is $D^*D$. This quantity is very sensitive to the mass of this near-threshold $cc\bar{u}\bar{d}$ state. For example, if we change the mass to 3876.0 (3880.0) MeV, the width would become 3.0 (9.7) MeV. The sensitivity is a general feature for near-threshold states. If we reduce the $cc\bar{u}\bar{d}$ mass to the central value of the LHCb $T_{cc}$, $M=3874.82$ MeV, the rearrangement decay width cannot be calculated with Eq. \eqref{decay} anymore, but one may consider the quasi two-body decay with
\cite{Capstick:1993kb,Roberts:1997kq,Segovia:2009zz,Ferretti:2014xqa,Gui:2018rvv},
\begin{gather}\label{threebody}
\Gamma=\int_0^{k_{max}}dk\frac{\Gamma_{D^{*+}\to D^0\pi^+}}{(M_{T_{cc}^+}-E_{D^{*+}}(k)-E_{D^{0}}(k))^2+\frac14\Gamma^2_{D^{*+}}}\frac{k^2|{\cal M}|^2}{2(2\pi)^2M_{T_{cc}^+}E_{D^{*+}}(k)E_{D^{0}}(k)},
\end{gather}
by noticing that $D^{*+}$ can decay into $D^0\pi^+$. In this formula, $\cal M$ is the decay amplitude of Eq. \eqref{amplitude}. $\Gamma_{D^{*+}\to D^0\pi^+}$ and $\Gamma_{D^{*+}}$ are the partial width and total width of $D^{*+}$ from the particle data book \cite{ParticleDataGroup:2022pth}, respectively. The $k_{max}$ is the maximum relative momentum for the $D^{*+}D^0$ system allowed by the three-body decay $D^{*+}\to (D^0\pi^+)D^0$ and is given by
\begin{gather}
k_{max}=\frac{\sqrt{M^2_{T_{cc}^+}-(2M_{D^0}+M_{\pi})^2}\sqrt{M^2_{T_{cc}^+}-M_{\pi}^2}}{2M_{T_{cc}^+}}.
\end{gather}
Using Eq. \eqref{threebody}, we obtain the width of the ground $cc\bar{u}\bar{d}$ state, $\Gamma\sim105$ keV. If we change the value of $\mathcal{C}$ to 6000 (8000) MeV, one gets $\Gamma\sim$68 (122) keV. The order of width is consistent with the measurement. At present, considering the model uncertainties, we conclude that the mass and decay width of the LHCb $T_{cc}$ may be qualitatively understood in the compact $cc\bar{u}\bar{d}$ tetraquark picture. This conclusion is consistent with that from the study on its production \cite{Qin:2020zlg,Jin:2021cxj}.

From Fig. \ref{ccqq-picture} and table \ref{decay1}, higher $cc\bar{n}\bar{n}$ states are all unstable. As for the high-mass $0(1^+)$ tetraquark located around 4074 MeV, the $D^*D^*$ (note $D^*D^*$ means both $D^{*+}D^{*+}$ and $\bar{D}^{*0}\bar{D}^{*0}$) channel is also open. The squared amplitude for the $D^*D$ channel is smaller than that for the $D^*D^*$ channel, but the phase space is bigger. As a result, the partial widths of these two channels are almost equal. The high-mass $1(0^+)$ state has a dominant decay channel $D^*D^*$ and a suppressed decay channel $DD$. The low-mass $1(0^+)$ state mainly decays into $DD$ since the $D^*D^*$ channel is kinematically forbidden. Our calculation shows that the $1(1^+)$ state is the broadest $cc\bar{n}\bar{n}$ state with $\Gamma$=53.0 MeV. This tetraquark and the high-mass $0(1^{+})$ state are almost degenerate, but one may distinguish them with the $D^*D^*$ channel. As shown in table \ref{decay1}, the $1(2^+)$ state is the second narrowest $cc\bar{n}\bar{n}$. It has only one rearrangement decay mode $D^*D^*$.\\

We now move on to the $cc\bar{n}\bar{s}$ system. The $K$ factors and rearrangement decay properties are also collected in tables \ref{Kij-ccqq} and \ref{decay1}, respectively. Compared to the low-mass $0^+$ state, the effective attraction between the $n$ and $s$ quarks in the lowest $1^+$ state makes its mass smaller. Fig. \ref{ccqq-picture}(b) tells us that all the tetraquarks are unstable. The low-mass $0^+$ state dominantly decays into $DD_s$. The high-mass $0^{+}$ state decays mainly into $D^*D_s^*$, while the $DD_s$ channel is suppressed because of the weak coupling. The lowest and second lowest $1^+$ states have two main rearrangement decay channels $D^*D_s$ and $DD^*_s$. Although the couplings with these two channels for the latter state are weaker than the former one, the bigger phase spaces induce larger widths.
For both tetraquarks, the ratio between the partial widths is around $\Gamma(D^*D_s):\Gamma(DD^*_s)\simeq1:1$. The highest $1^+$ $cc\bar{n}\bar{s}$ mainly decays into $D^*D^*_s$. The ratio between its different partial widths is
\begin{gather}
\Gamma(D^*D^*_s):\Gamma(D^*D_s):\Gamma(DD^*_s)\simeq 6.7:1.0:1.1 .
\end{gather}
For the tensor state, there is only one decay channel $D^*D^*_s$.

As for the $cc\bar{s}\bar{s}$ case, the features in the amplitude $\mathcal{M}$ and spectrum are similar to the isovector $cc\bar{n}\bar{n}$ case. However, The widths of $cc\bar{s}\bar{s}$ states (table \ref{decay1}) are narrower than $cc\bar{n}\bar{n}$ because of smaller phase spaces. The lowest and highest $0^+$ states mainly decay into $D_sD_s$ and $D^*_sD^*_s$, respectively. The broadest $cc\bar{s}\bar{s}$ tetraquark is the $1^+$ state with $\Gamma\sim42.7$ MeV, which only decays into $D^*_sD_s$. The $2^+$ state has the smallest width $\Gamma\sim14.6$ MeV. Although the search of $X_{cc\bar{s}\bar{s}}$ in the invariant mass spectrum of $D_s^+D_s^+$ and $D_s^{*+}D_s^{*+}$ by Belle \cite{Belle:2021kub} gives a negative result, theoretical analysis still favors the possibility for the existence of $cc\bar{s}\bar{s}$ resonances.

\subsection{$bb\bar{n}\bar{n}$, $bb\bar{n}\bar{s}$, and $bb\bar{s}\bar{s}$ systems}

\begin{figure}[htbp]
	\centering
	\begin{minipage}[b]{0.32\textwidth}
		\centering
		\includegraphics[width=1\textwidth]{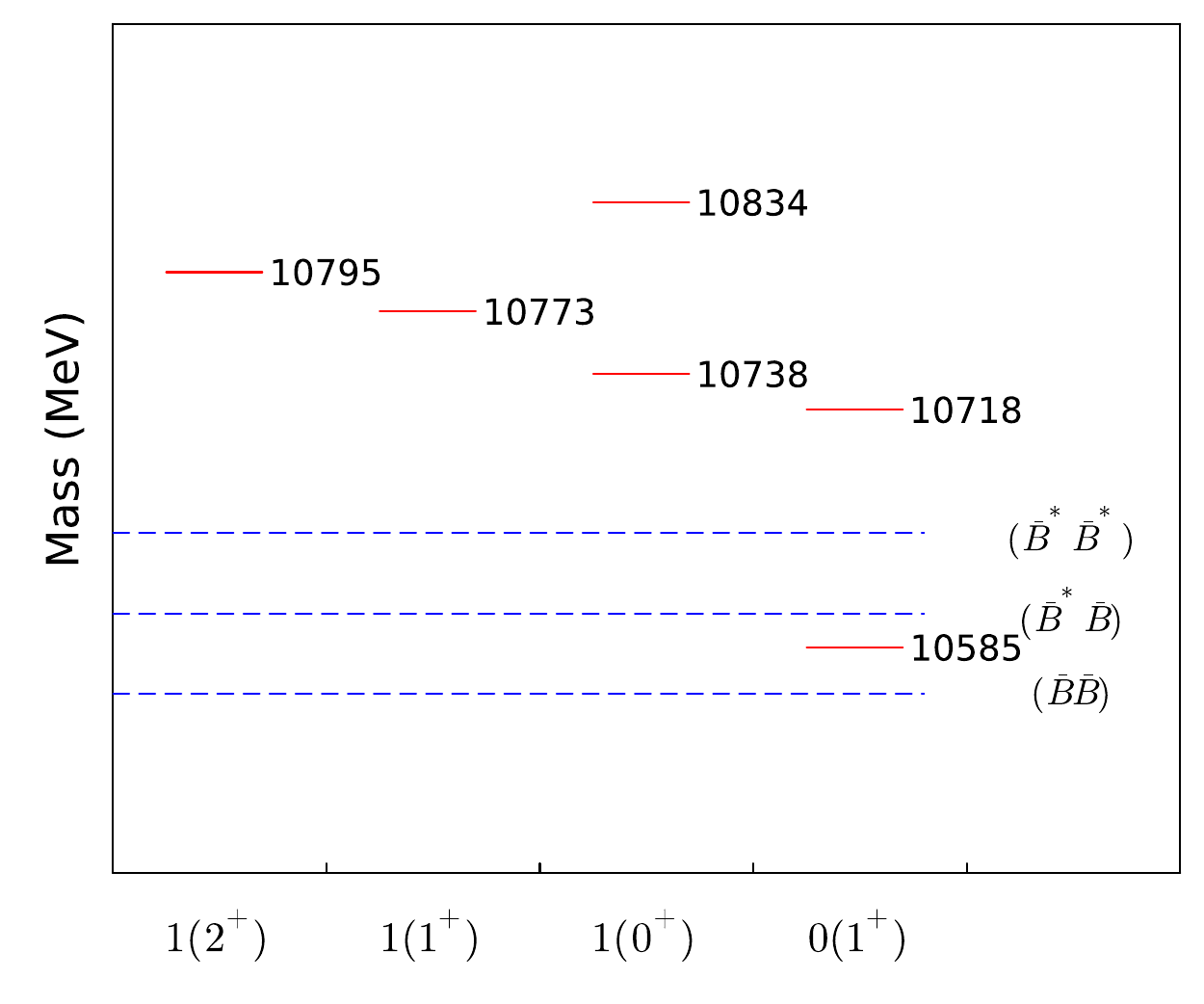}
		\subcaption{$bb\bar{n}\bar{n}$}
	\end{minipage}
	\begin{minipage}[b]{0.32\textwidth}
		\centering
		\includegraphics[width=1\textwidth]{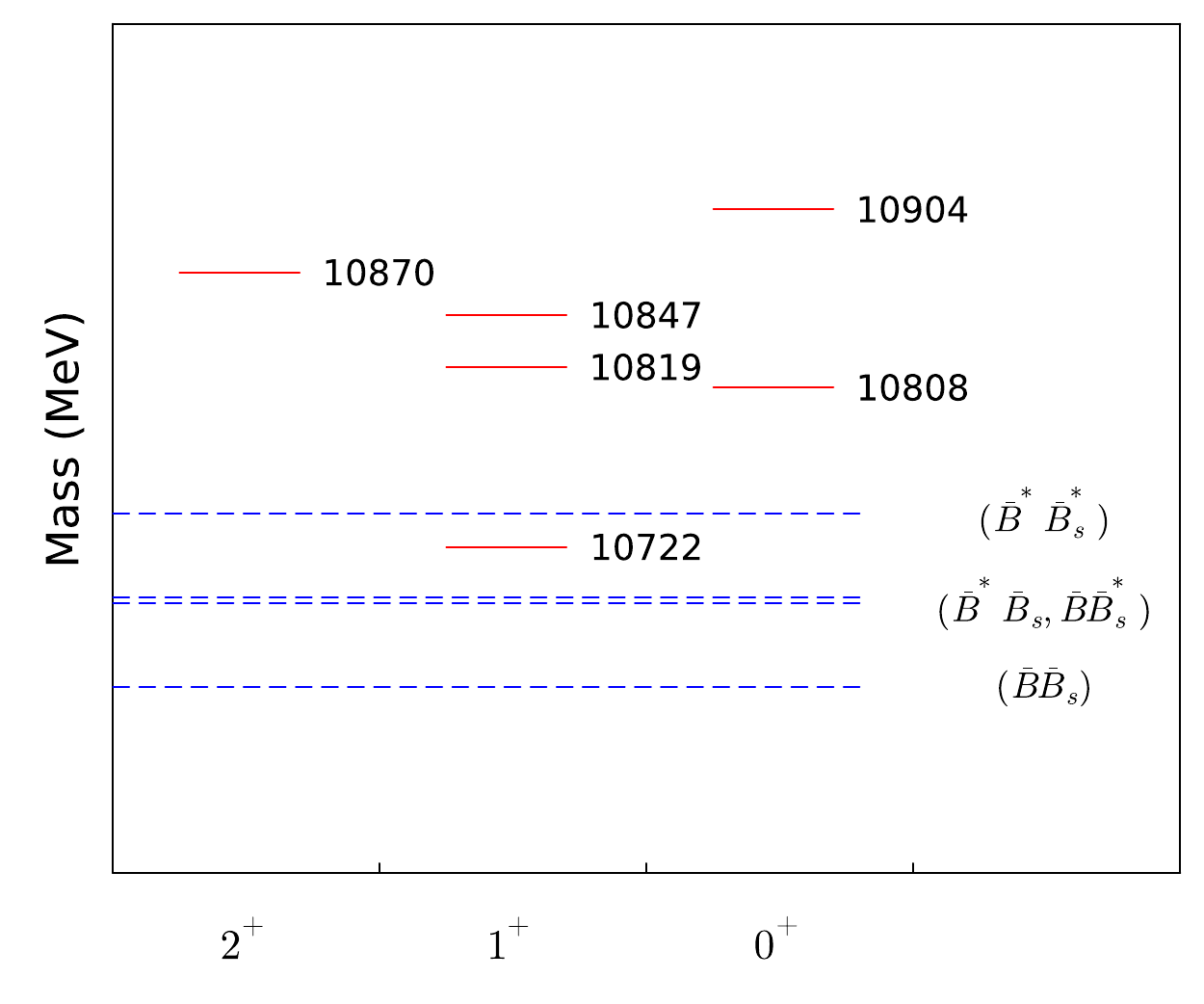}
		\subcaption{$bb\bar{n}\bar{s}$}
	\end{minipage}
	\begin{minipage}[b]{0.32\textwidth}
		\centering
		\includegraphics[width=1\textwidth]{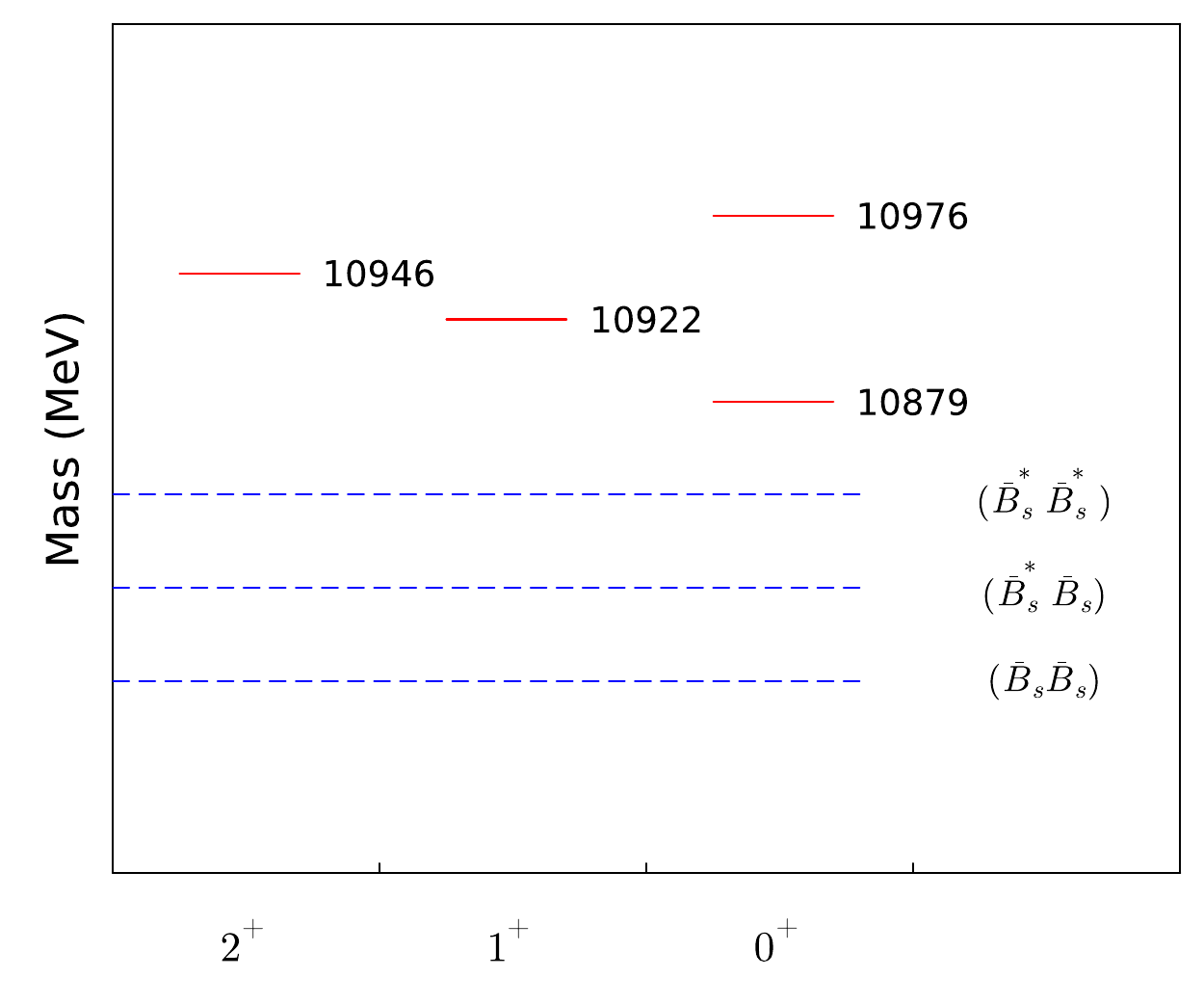}
		\subcaption{$bb\bar{s}\bar{s}$}
	\end{minipage}
	\caption{Relative positions for the considered tetraquark states. The red solid lines represent the estimated masses with the mass splitting scheme. The blue dashed lines indicate various meson-meson thresholds. \label{bbqq-picture}}
\end{figure}

We show the mass results for the $bb\bar{q}\bar{q}$ states in table \ref{m2} and their relative positions in Fig. \ref{bbqq-picture}. The information about effective chromomagnetic interactions and decay properties in the rearrangement mechanism is placed in tables \ref{Kij-ccqq} and \ref{decay1}, respectively.

\begin{table}[!h]\scriptsize\centering
	\caption{Mass spectrum for the $bb\bar{q}\bar{q}$ states in units of MeV. The last three columns list the results obtained with Eqs. \eqref{mass}, \eqref{mref}, and \eqref{mCMI}, respectively. The thresholds of $\bar{B}\bar{B}$, $\bar{B}\bar{B}_s$, and $\bar{B}_s\bar{B}_s$ are used in getting the lower limits for the tetraquark masses in the cases of $bb\bar{n}\bar{n}$, $bb\bar{n}\bar{s}$, and $bb\bar{s}\bar{s}$, respectively. The upper limits are obtained using Eq\eqref{mCMI}, where the effective quark masses  are $m_n=361.8$ MeV, $m_s=542.4$ MeV, $m_c=1724.1$ MeV, and $m_b=5054.4$ MeV, respectively. We take values in the mass splitting scheme (fourth column) in this article.}\scriptsize\label{m2}
	\begin{tabular}{c|cccccc}\hline
		\hline\multicolumn{6}{c}{$bb\bar{n}\bar{n}$ system} \\\hline\hline
		$I(J^{P})$ & $\langle H_{CMI} \rangle$ & $E_{CMI}$ &Mass&Lower limits&Upper limits\\\hline
		$1(2^{+})$ &$\left(\begin{array}{c}65.1\end{array}\right)$&$\left(\begin{array}{c}65.1\end{array}\right)$&$\left(\begin{array}{c}10795.3\end{array}\right)$&$\left(\begin{array}{c}10691.1\end{array}\right)$&$\left(\begin{array}{c}10897.5\end{array}\right)$\\
		$1(1^{+})$ &$\left(\begin{array}{c}42.7\end{array}\right)$&$\left(\begin{array}{c}42.7\end{array}\right)$&$\left(\begin{array}{c}10772.9\end{array}\right)$&$\left(\begin{array}{c}10668.7\end{array}\right)$&$\left(\begin{array}{c}10875.1\end{array}\right)$\\
		$1(0^{+})$ &$\left(\begin{array}{cc}31.5&41.2\\41.2&80.8\end{array}\right)$&$\left(\begin{array}{c}104.1\\8.2\end{array}\right)$&$\left(\begin{array}{c}10834.4\\10738.4\end{array}\right)$&$\left(\begin{array}{c}10730.1\\10634.2\end{array}\right)$&$\left(\begin{array}{c}10936.5\\10840.6\end{array}\right)$\\
		$0(1^{+})$ &$\left(\begin{array}{cc}-141.3&-23.8\\-23.8&-16.8\end{array}\right)$&$\left(\begin{array}{c}-12.4\\-145.7\end{array}\right)$&$\left(\begin{array}{c}10717.8\\10584.5\end{array}\right)$&$\left(\begin{array}{c}10613.6\\10480.3\end{array}\right)$&$\left(\begin{array}{c}10820.0\\10686.7\end{array}\right)$\\
		\hline\multicolumn{6}{c}{$bb\bar{n}\bar{s}$ system} \\\hline\hline
		$J^{P}$ & $\langle H_{CMI} \rangle$ & $E_{CMI}$ &Mass&Lower limits&Upper limits\\\hline
		$2^{+}$ &$\left(\begin{array}{c}49.1\end{array}\right)$&$\left(\begin{array}{c}49.1\end{array}\right)$&$\left(\begin{array}{c}10869.9\end{array}\right)$&$\left(\begin{array}{c}10765.7\end{array}\right)$&$\left(\begin{array}{c}11062.1\end{array}\right)$\\
		$1^{+}$ &$\left(\begin{array}{ccc}25.6&-0.8&-1.6\\-0.8&-91.7&-24.9\\-1.6&-24.9&-8.5\end{array}\right)$&$\left(\begin{array}{c}25.7\\-1.7\\-98.6\end{array}\right)$&$\left(\begin{array}{c}10846.5\\10819.1\\10722.2\end{array}\right)$&$\left(\begin{array}{c}10742.3\\10714.9\\10618.0\end{array}\right)$&$\left(\begin{array}{c}11038.7\\11011.3\\10914.4\end{array}\right)$\\
		$0^{+}$ &$\left(\begin{array}{cc}13.9&43.1\\43.1&56.0\end{array}\right)$&$\left(\begin{array}{c}82.9\\-13.0\end{array}\right)$&$\left(\begin{array}{c}10903.8\\10807.8\end{array}\right)$&$\left(\begin{array}{c}10799.5\\10703.6\end{array}\right)$&$\left(\begin{array}{c}11095.9\\11000.0\end{array}\right)$\\
		\hline\multicolumn{6}{c}{$bb\bar{s}\bar{s}$ system} \\\hline\hline
		$J^{P}$ & $\langle H_{CMI} \rangle$ & $E_{CMI}$ &Mass&Lower limits&Upper limits\\\hline
		$2^{+}$ &$\left(\begin{array}{c}34.7\end{array}\right)$&$\left(\begin{array}{c}34.7\end{array}\right)$&$\left(\begin{array}{c}10946.1\end{array}\right)$&$\left(\begin{array}{c}10841.9\end{array}\right)$&$\left(\begin{array}{c}11228.3\end{array}\right)$\\
		$1^{+}$ &$\left(\begin{array}{c}10.1\end{array}\right)$&$\left(\begin{array}{c}10.1\end{array}\right)$&$\left(\begin{array}{c}10921.6\end{array}\right)$&$\left(\begin{array}{c}10817.3\end{array}\right)$&$\left(\begin{array}{c}11203.7\end{array}\right)$\\
		$0^{+}$ &$\left(\begin{array}{cc}-2.1&45.1\\45.1&33.6\end{array}\right)$&$\left(\begin{array}{c}64.2\\-32.7\end{array}\right)$&$\left(\begin{array}{c}10975.7\\10878.7\end{array}\right)$&$\left(\begin{array}{c}10871.4\\10774.5\end{array}\right)$&$\left(\begin{array}{c}11257.8\\11160.9\end{array}\right)$\\
		\hline
	\end{tabular}
\end{table}

For the $bb\bar{n}\bar{n}$ system, the feature in mass spectrum is similar to the $cc\bar{n}\bar{n}$ case. From table \ref{m2}, the lowest isoscalar $1^+$ state is $\sim$20 MeV below the $\bar{B}^*\bar{B}$ threshold (10604 MeV). It indicates that this tetraquark is stable against strong decay. Almost all theoretical studies favor this conclusion \cite{Liu:2019zoy,Cheng:2020wxa,Zhang:2007mu,Ren:2023pip}. In Ref. \cite{Lu:2020rog}, the authors investigated the $bb\bar{n}\bar{n}$ states using a relativized quark model. Their results indicate that the lowest $I(J^P)=0(1^+)$ $bb\bar{u}\bar{d}$ state is 54 MeV below the relevant $\bar{B}\bar{B}^*$ threshold. In a recent study \cite{Kim:2022mpa}, the authors also found that a ground $bb\bar{u}\bar{d}$ tetraquark is stable with the binding energy 115 MeV in a quark-quark-antidiquark picture. Except this lowest tetraquark, other $bb\bar{n}\bar{n}$ states have rearrangement decay modes. The ratio between its two partial widths for the high-mass $0(1^+)$ tetraquark is $\Gamma(\bar{B}^*\bar{B}^*):\Gamma(\bar{B}^*\bar{B})\simeq1:1.5$ while the low-mass (high-mass) $1(0^+)$ state dominantly decays into $\bar{B}\bar{B}$ ($\bar{B}^*\bar{B}^*$). In the assumption that $bb\bar{n}\bar{n}$ and $cc\bar{n}\bar{n}$ systems involve the same $\mathcal{C}$, from table \ref{decay1}, the estimated decay width of a $bb\bar{n}\bar{n}$ state is narrower than the corresponding $cc\bar{n}\bar{n}$. It is due to the smaller phase space in the double-bottom case. The qualitative feature of width is in accordance with the conclusion that the doubly heavy tetraquark system would have stable state(s) when the heavy quark mass increases \cite{Manohar:1992nd,Michael:1999nq,Vijande:2009kj,Du:2012wp}.

For the $bb\bar{n}\bar{s}$ states, the feature of mass spectrum is similar to the $cc\bar{n}\bar{s}$ case. Their rearrangement decay widths are narrower than the corresponding $cc\bar{n}\bar{s}$ states when one adopts the same $\mathcal{C}$, but the feature of dominant decay channels may be different. From table \ref{decay1}, we can find that the second lowest $1^+$ $bb\bar{n}\bar{s}$ state decays into $\bar{B}^*\bar{B}_s^*$, $\bar{B}^*\bar{B}_s$, and $\bar{B}\bar{B}_s^*$. The ratio between partial widths of these channels is $\Gamma(\bar{B}^*\bar{B}_s^*): \Gamma(\bar{B}^*\bar{B}_s):\Gamma(\bar{B}\bar{B}_s^*)\simeq 4:1:1.3$. The first channel is its main decay mode, but the latter two  corresponding channels in the $cc\bar{n}\bar{s}$ case are the main decay modes. The highest $1^+$ $bb\bar{n}\bar{s}$ mainly decays into $\bar{B}^*\bar{B}_s$ and $\bar{B}\bar{B}_s^*$ while its charm analog mainly decays into two vector mesons. For the other $bb\bar{n}\bar{s}$ states, similar decay properties to their charm analogs can be seen. 

The spectrum and decay properties in the $bb\bar{s}\bar{s}$ case have similar features to the $cc\bar{s}\bar{s}$ case. No stable state is observed. Comparing $bb\bar{s}\bar{s}$ and $cc\bar{s}\bar{s}$ results in table \ref{decay1}, one also finds that the decay width becomes narrower with the increasing heavy quark mass when the same $\mathcal{C}$ is used. The high-mass $0^+$ tetraquark dominantly decays into $\bar{B}^*_s\bar{B}^*_s$, while the low-mass $0^+$ mainly decays into $\bar{B}_s\bar{B}_s$. Both $1^+$ and $2^+$ $bb\bar{s}\bar{s}$ states have only one decay channel, $\bar{B}^*_s\bar{B}_s$ for the former state and $\bar{B}^*_s\bar{B}^*_s$ for the latter one.

\subsection{$bc\bar{n}\bar{n}$, $bc\bar{s}\bar{s}$, and $bc\bar{n}\bar{s}$ systems}

For the $bc\bar{n}\bar{n}$ and $bc\bar{n}\bar{s}$ systems, we have explored their mass spectra, $K$ factors, and rearrangement decay properties in Ref. \cite{Cheng:2020nho} with the mass splitting scheme. The numerical results may be slightly different because of the adopted parameters, but the difference has no effect on the conclusions. We do not show the results here again.

For the $bc\bar{n}\bar{n}$ system, all the obtained tetraquark states are above corresponding meson-meson thresholds. They will dissociate into a pair of heavy-light mesons through the rearrangement decay mechanism. This observation is consistent with Ref. \cite{Cheng:2020wxa} where the heavy diquark-antiquark symmetry is adopted. It is worth mentioning that the mass of the lowest $I(J^P)=0(0^+)$ $bc\bar{n}\bar{n}$ is predicted to be 7151.5 MeV which is very close to the $\bar{B}D$ threshold (7145 MeV). In Ref. \cite{Karliner:2017qjm}, Karliner and Rosner got a mass 11 MeV below the $\bar{B}^0D^0$ threshold for this state. However, it is not sufficient to conclude whether this tetraquark is actually above or below the $\bar{B}D$ threshold because of the precision of their result. Our lowest $0(1^+)$ state is also slightly above the relevant $\bar{B}^*D$ threshold. In a recent lattice QCD study \cite{Alexandrou:2023cqg}, evidence for  $\bar{b}\bar{c}ud$ tetraquark shallow bound states in both $J=0$ and $J=1$ cases is observed. From theoretical calculations, it is very possible that exotic states around both $\bar{B}D$ and $\bar{B}^*D$ thresholds exist. For discussions on other $bc\bar{n}\bar{n}$ and $bc\bar{n}\bar{s}$ states, one may consult Ref. \cite{Cheng:2020nho}. The ratio between different partial widths of a state may be obtained with the results in table VI or VIII of that paper.

\setlength{\tabcolsep}{0.3mm}
\begin{table}[htbp]\centering\scriptsize
	\caption{Mass spectrum for the $bc\bar{s}\bar{s}$ states in units of MeV. The last three columns list the results obtained with Eqs. \eqref{mass}, \eqref{mref}, and \eqref{mCMI}, respectively. The threshold of $D_s\bar{B}_s$ is used in getting the lower limits for the tetraquark masses.  The upper limits are obtained using Eq\eqref{mCMI}, where the effective quark masses  are $m_n=361.8$ MeV, $m_s=542.4$ MeV, $m_c=1724.1$ MeV, and $m_b=5054.4$ MeV, respectively.
 We take values in the mass splitting scheme (fourth column) in this article.}\scriptsize\label{m3}
	\begin{tabular}{c|cccccc}\hline
		\hline\multicolumn{6}{c}{$bc\bar{s}\bar{s}$ system} \\\hline\hline
		$J^{P}$ & $\langle H_{CMI} \rangle$ & $E_{CMI}$ &Mass&Lower limits&Upper limits\\\hline
		$2^{+}$ &$\left(\begin{array}{c}46.7\end{array}\right)$&$\left(\begin{array}{c}46.7\end{array}\right)$&$\left(\begin{array}{c}7617.9\end{array}\right)$&$\left(\begin{array}{c}7525.8\end{array}\right)$&$\left(\begin{array}{c}7910.0\end{array}\right)$\\
		$1^{+}$ &$\left(\begin{array}{ccc}-1.3&35.2&16.6\\35.2&23.3&-50.9\\16.6&-50.9&1.3\end{array}\right)$&$\left(\begin{array}{c}69.3\\15.2\\-61.1\end{array}\right)$&$\left(\begin{array}{c}7640.5\\7586.4\\7510.2\end{array}\right)$&$\left(\begin{array}{c}7548.4\\7494.3\\7418.0\end{array}\right)$&$\left(\begin{array}{c}7932.6\\7878.5\\7802.2\end{array}\right)$\\
		$0^{+}$ &$\left(\begin{array}{cc}-25.3&88.2\\88.2&34.0\end{array}\right)$&$\left(\begin{array}{c}97.4\\-88.7\end{array}\right)$&$\left(\begin{array}{c}7668.6\\7482.6\end{array}\right)$&$\left(\begin{array}{c}7576.5\\7390.4\end{array}\right)$&$\left(\begin{array}{c}7960.7\\7774.6\end{array}\right)$\\
		\hline
	\end{tabular}
\end{table}

\begin{figure}[htbp]
	\centering
	\begin{minipage}{0.32\textwidth}		\centering
		\includegraphics[width=1\textwidth]{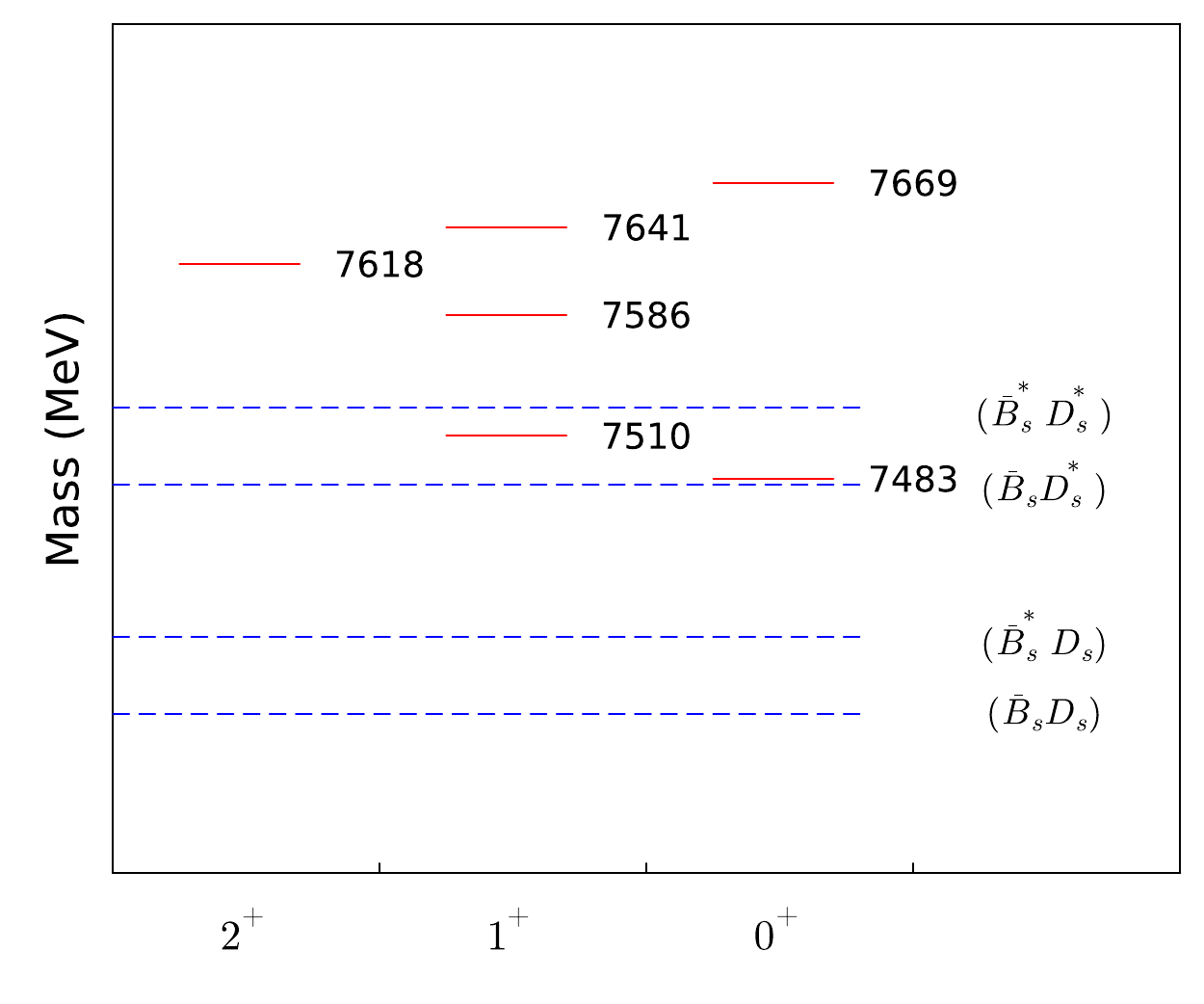}
		\subcaption{$bc\bar{s}\bar{s}$}
	\end{minipage}
	\caption{Relative positions for the $bc\bar{s}\bar{s}$ states. The red solid lines represent the estimated masses with the mass splitting scheme. The blue dashed lines indicate various meson-meson thresholds. \label{bcqq-picture}}
\end{figure}

\setlength{\tabcolsep}{0.1mm}\begin{table}[htbp]\caption{$K_{ij}$'s for the $bc\bar{s}\bar{s}$ case.}\label{Kij-bcqq}\scriptsize\centering
	\begin{tabular}{ccccccc}\hline\hline
        &\multicolumn{6}{c}{$bc\bar{s}\bar{s}$}\\
		\hline
		$J^{P}$&$K_{bc}$&$K_{b\bar{s}}$&$K_{c\bar{s}}$&$K_{ss}$&&\\
		$2^{+}$&$\left[\begin{array}{c}2.7\end{array}\right]$&$\left[\begin{array}{c}2.7\end{array}\right]$&$\left[\begin{array}{c}2.7\end{array}\right]$&$\left[\begin{array}{c}2.7\end{array}\right]$&&\\
		$1^{+}$&$\left[\begin{array}{c}-2.9\\-0.9\\-2.8\end{array}\right]$&$\left[\begin{array}{c}2.2\\-7.4\\2.5\end{array}\right]$&$\left[\begin{array}{c}7.0\\2.5\\-12.2\end{array}\right]$&$\left[\begin{array}{c}3.5\\2.7\\3.1\end{array}\right]$&&\\
		$0^{+}$&$\left[\begin{array}{c}3.5\\3.1\end{array}\right]$&$\left[\begin{array}{c}7.4\\-12.8\end{array}\right]$&$\left[\begin{array}{c}7.5\\-12.8\end{array}\right]$&$\left[\begin{array}{c}3.5\\3.1\end{array}\right]$&&\\
		\hline\hline
	\end{tabular}
\end{table}

\begin{table}[htbp]\scriptsize\centering
	\begin{tabular}{c|c|cccc}\hline\hline
		$J^{P}$&Mass&\multicolumn{3}{c}{Channels}&$\Gamma$\\
		\hline
		&&$\bar{B}_s^{*} D_s^*$&&\\
		$2^{+}$&$\left[\begin{array}{c}7617.9\end{array}\right]$&$\left[\begin{array}{c}(33.3, 12.8)\end{array}\right]$&&&$\left[\begin{array}{c}12.8\end{array}\right]$\\
		&&$\bar{B}_s^*D_s^*  $&$\bar{B}_s^*D_s$&$\bar{B}_sD^*_s$&\\
		$1^{+}$&$\left[\begin{array}{c}7640.5\\7586.4\\7510.2 \end{array}\right]$&$\left[\begin{array}{c}(46.2, 19.7)\\(3.0, 0.9)\\(0.8, -) \end{array}\right]$&$\left[\begin{array}{c}(0.4, 0.2)\\(1.4, 0.8)\\(39.8, 18.2) \end{array}\right]$&$\left[\begin{array}{c}(8.1, 4.1)\\(29.3, 12.3)\\(4.3, 1.0) \end{array}\right]$&$\left[\begin{array}{c}24.1\\14.1\\19.2\end{array}\right]$\\
		&&$\bar{B}_s^*D_s^*$&$\bar{B}_sD_s$&&\\
		$0^{+}$&$\left[\begin{array}{c}7668.6\\7482.6\end{array}\right]$&$\left[\begin{array}{c}(55.2, 26.1)\\(3.2, -)\end{array}\right]$&$\left[\begin{array}{c}(0.2, 0.1)\\(41.5, 20.6)\end{array}\right]$&&$\left[\begin{array}{c}26.3\\20.6\end{array}\right]$\\
		\hline\hline
	\end{tabular}\caption{Rearrangement decays for the $bc\bar{s}\bar{s}$ case. The two numbers in the parenthesis for a decay channel mean dimensionless $100|\mathcal{M}|^2/{\mathcal C}^2$ and dimensional partial width, respectively. The masses and widths are presented in units of MeV.}\label{decay4}
\end{table}

The numerical results for the mass spectrum and $K$ factors of the $bc\bar{s}\bar{s}$ system are given in tables \ref{m3} and \ref{Kij-bcqq}, respectively. The relative positions of the states are illustrated in Fig. \ref{bcqq-picture} and their rearrangement decay properties are listed in table \ref{decay4}. There are six possible tetraquark states in the $bc\bar{s}\bar{s}$ system. As shown in table \ref{decay4}, their decay widths are not broad, ranging from 12 MeV to 26 MeV. The two $0^+$ states have different dominant decay channels although their widths are almost the same. The lowest $1^+$ state has two decay channels, but it mainly decays into $\bar{B}_s^*D_s$. The ratio of partial widths for the second lowest $1^+$ state is estimated to be $\Gamma(\bar{B}^*_s D_s^*): \Gamma(\bar{B}^*_s D_s):\Gamma(\bar{B}_s D_s^*)\simeq 1.2:1.0:15.4 $. For the highest $1^+$ state, the ratio of partial widths between $\bar{B}^*_s D_s^*$ and $\bar{B}_s D_s^*$ channels is about $\Gamma(\bar{B}^*_s D_s^*):\Gamma(\bar{B}_s D_s^*)\simeq 4.8:1.0$. The narrowest $bc\bar{s}\bar{s}$ tetraquark has $J^P=2^+$ with $\Gamma\sim12$ MeV and has only one rearrangement decay channel $\bar{B}^*_s D_s^*$.

\section{Summary and discussions}\label{sec4}

In this work, we have studied systematically the mass spectrum and rearrangement decays for the $QQ\bar{q}\bar{q}$ tetraquark states in the framework of a mass splitting model. The masses are estimated by using the reference state $X(4140)$ with the assumption that it is the lowest $1^{++}$ $cs\bar{c}\bar{s}$ tetraquark. The decay widths for the $QQ\bar{q}\bar{q}$ states above the relevant meson-meson thresholds are described by a simple rearrangement scheme. Based on our mass and width results as well as the production result for $T_{cc}$ \cite{Jin:2021cxj}, one finds:

(i) The lowest $0(1^+)$ $cc\bar{u}\bar{d}$ tetraquark state can be used to understand the LHCb $T_{cc}$ state.

(ii) In the $S$-wave $QQ\bar{q}\bar{q}$ states, only the $0(1^+)$ $bb\bar{u}\bar{d}$ tetraquark is stable. It is located $\sim$20 MeV below the $\bar{B}\bar{B}$ threshold.

(iii) The lowest $0(0^+)$ $bc\bar{u}\bar{d}$ state has a mass about 7152 MeV and is very close to the $\bar{B}D$ threshold. We can not exclude it as a stable tetraquark because of the adopted model. The lowest $0(1^+)$ $bc\bar{u}\bar{d}$ is also a near-threshold state.

In our mass splitting model, the mass formulas for $QQ\bar{q}\bar{q}$ and $Q\bar{Q}q\bar{q}$ tetraquarks have the same form. The mass differences between these states are mainly determined by the eigenvalues of $H_{CMI}$. We here take a look at two tetraquark states related to the observed $X(3872)$ and $T_{cc}(3875)$. The former state is the low-mass $I(J^{PC})=0(1^{++})$ $c\bar{c}n\bar{n}$ \cite{Wu:2018xdi} and the latter the low-mass $I(J^P)=0(1^+)$ $cc\bar{n}\bar{n}$. From the values of $E_{CMI}$, the ground $0(1^{++})$ $c\bar{c}n\bar{n}$ is about (-97)-(-172)=75 MeV heavier than the ground $0(1^+)$ $cc\bar{n}\bar{n}$. Since the observed $X(3872)$ and $T_{cc}(3875)$ have almost the same mass, one cannot interpret the $X(3872)$ as a compact tetraquark without additional assumption if the LHCb $T_{cc}$ is really a compact tetraquark.

The tetraquark spectrum would be shifted downward or upward with some scale once a different reference meson is adopted. We took the $X(4140)$ as the reference tetraquark in the above discussions. One may also use the mass of $X(4274)$ as the reference scale by assuming it to be the high-mass $1^{++}$ $cs\bar{c}\bar{s}$ state. In that case, one finds that all the $QQ\bar{q}\bar{q}$ masses would be $\sim$27 MeV smaller. Then, the following four lowest states would be stable against strong decay: $0(1^+)$ $cc\bar{n}\bar{n}$ around 3851 MeV, $0(1^+)$ $bb\bar{n}\bar{n}$ around 10559 MeV, $0(1^+)$ $bc\bar{n}\bar{n}$ around 7181 MeV, and $0(0^+)$ $bc\bar{n}\bar{n}$ around 7125 MeV. The $X(3872)$ is a well-established exotic meson, but its nature, a $D\bar{D}^*$ molecule, a $c\bar{c}n\bar{n}$ compact tetraquark, or a charmonium state affected strongly by coupled channels, is still under discussions. If one treats it as a compact tetraquark and takes it rather than $X(4140)$ as the reference, all the studied states would become $M_{T_{cc}}+75\,\mathrm{MeV}-M_{X(3872)}\sim$78 MeV lighter and more stable tetraquarks are obtained. In this case, the observed $T_{cc}(3875)$ is more likely a molecule than a compact tetraquark and we cannot consistently interpret the $X(4140)$ to be a compact tetraquark.

Since the model parameters have uncertainties, interpreting the LHCb $T_{cc}$ as a compact tetraquark needs more investigations. The mass for the lowest $0(1^+)$ $cc\bar{u}\bar{d}$ tetraquark in the present method is about 50 MeV smaller than the analysis using the heavy diquark-antiquark symmetry \cite{Cheng:2022qcm}. It is difficult to judge which method is more reliable. If a compact tetraquark should be $\sim$25 MeV above the $DD^*$ threshold, it is also possible that the LHCb $T_{cc}$ is a tetraquark state affected by the $DD^*$ interaction.

Due to limitation of measurements, we take a universal coupling parameter $\mathcal{C}$ for different systems. The results on decay widths should differ from future measured values, but the width ratios between different channels (even different states) calculated by eliminating $\mathcal{C}$ with our results may be used to check the compact picture. Although the adopted scheme is crude, the study on width highlights dominant decay channels for the doubly heavy tetraquark states.

\section*{Acknowledgments}

We are grateful for the helpful discussions with Dr. Yu-Nan Liu. This project was supported by the National Natural Science Foundation of China (Nos. 12235008, 12275157, 12475143, and 11905114) and the Shandong Province Natural Science Foundation (ZR2023MA041).

\end{document}